\documentclass[12pt]{article}

\usepackage{graphicx,amssymb,amsmath,dsfont}
\usepackage{bbm} 

\usepackage[a4paper]{geometry} 
\def\dfrac#1#2{\frac{\displaystyle\strut #1}{\displaystyle\strut #2}}

\def\bra#1{\mathinner{\langle{#1}|}}
\def\ket#1{\mathinner{|{#1}\rangle}}
\def\braket#1{\mathinner{\langle{#1}\rangle}}

\begin{document}

\title{\LARGE\bf 
Optimized entanglement for quantum parameter estimation
from noisy qubits
}

\author{Fran\c{c}ois {\sc Chapeau-Blondeau}, \\
    Laboratoire Angevin de Recherche en Ing\'enierie des Syst\`emes (LARIS), \\
    Universit\'e d'Angers,
    62 avenue Notre Dame du Lac, 49000 Angers, France.
}

\date{\today}

\maketitle

\parindent=8mm \parskip=0ex

\begin{abstract}
For parameter estimation from an $N$-component composite quantum system, it is known that a 
separable preparation leads to a mean-squared estimation error scaling as $1/N$ while an 
entangled preparation can in some conditions afford a smaller error with $1/N^2$ scaling. 
This quantum superefficiency is however very fragile to noise or decoherence, and typically 
disappears with any small amount of random noise asymptotically at large $N$. To complement this 
asymptotic characterization, here we characterize how the estimation efficiency evolves as a 
function of the size $N$ of the entangled system and its degree of entanglement. We address a 
generic situation of qubit phase estimation, also meaningful for frequency estimation. 
Decoherence is represented by the 
broad class of noises commuting with the phase rotation, which includes depolarizing, phase-flip, 
and thermal quantum noises. In these general conditions, explicit expressions are derived for the 
quantum Fisher information quantifying the ultimate achievable efficiency for estimation. We 
confront at any size $N$ the efficiency of the optimal separable preparation to that of an 
entangled preparation with arbitrary degree of entanglement. We exhibit the $1/N^2$ 
superefficiency with no noise, and prove its asymptotic disappearance at large $N$ for any 
non-vanishing noise configuration. For maximizing the estimation efficiency, we characterize the 
existence of an optimum $N_{\rm opt}$ of the size of the entangled system along with an optimal 
degree of entanglement. For nonunital noises, maximum efficiency is usually obtained at partial 
entanglement. 
Grouping the $N$ qubits into independent blocks formed of $N_{\rm opt}$ entangled qubits restores at 
large $N$ a nonvanishing efficiency that can improve over that of $N$ independent qubits optimally 
prepared. Also, one inactive qubit included in the entangled probe sometimes stands as the most 
efficient setting for estimation. The results further attest with new characterizations the 
subtlety of entanglement for quantum information in the presence of noise, showing that when 
entanglement is beneficial maximum efficiency is not necessarily obtained by maximum entanglement 
but instead by a controlled degree and finite optimal amount of it.
\end{abstract}

\section{Introduction}

{\let\thefootnote\relax\footnote{{Preprint of an article published in {\em International 
Journal of Quantum Information}, Vol.~16, No.~07 (2018) 1850056 (25 pages) 
DOI: 10.1142/S0219749918500569 \\
Copyright World Scientific Publishing Company www.worldscientific.com/worldscinet/ijqi }}}
Quantum information processing can exploit resources non-existing classically and offering 
unparalleled means for enhanced efficiency. This is the case with quantum entanglement, and 
in this paper we further investigate some important aspects of entanglement affording specific 
benefit for quantum estimation. Typically, in a task of quantum parameter estimation 
\cite{Helstrom76,Paris09,Brivio10,Chappell11,Smirne15,Chapeau15,Assad17}, 
one has access to a quantum system with a state $\rho_\xi$ carrying the dependence with the 
unknown parameter $\xi$ to be estimated. After measuring the quantum system in state $\rho_\xi$, 
the measurement outcomes are processed by means of an estimator $\widehat{\xi}$ to infer a value 
for the unknown parameter $\xi$. Following any measurement protocol, any conceivable estimator 
$\widehat{\xi}$ for $\xi$ is endowed with a mean-squared error 
$\langle (\widehat{\xi} -\xi)^2 \rangle$ which is lower bounded by the Cram\'er-Rao bound 
involving the reciprocal of the classical Fisher information $F_c(\xi)$ \cite{Cover91,Kay93}. 
Estimators are known, such as the maximum likelihood estimator, that 
can reach the Cram\'er-Rao bound in definite (usually asymptotic) conditions. 
Higher Fisher information $F_c(\xi)$ generally entails higher efficiency in estimation,
and one has then the faculty to select the measurement protocol so as to maximize $F_c(\xi)$.
In this respect, there is a fundamental upper bound \cite{Braunstein94,Paris09}
provided by the quantum Fisher information
$F_q(\xi)$ which sets a limit to the classical Fisher information $F_c(\xi)$, i.e.\ fixing
$F_c(\xi) \le F_q(\xi)$. In turn, constructive methodologies (usually adaptive) exist 
\cite{Barndorff00,Armen02,Fujiwara06,Brivio10,Tesio11,Okamoto12}
yielding a measurement protocol reaching $F_c(\xi) = F_q(\xi)$. The quantum Fisher 
information $F_q(\xi)$ has thus the status of a fundamental metric characterizing the ultimate 
best efficiency achievable in quantum estimation, and we shall use it as such in this report.

A definite measurement operating on a single copy of the quantum system in state $\rho_\xi$ 
can be repeated $N$ times on $N$ independent copies of the quantum system. In this situation 
of independent state preparation and independent measurement, the quantum Fisher information 
is additive \cite{Ji08} and amounts to $NF_q(\xi)$. This is associated with a mean-squared error 
$\langle (\widehat{\xi} -\xi)^2 \rangle$ evolving as $1/N$, forming the standard or shot-noise 
limit of the error \cite{Giovannetti04}. There however exists a possibility of improving over such 
$1/N$ efficiency, by exploiting the specifically quantum property of entanglement. With the 
$N$-component composite quantum system, exploiting entanglement for estimation can lead to
a smaller mean-squared error scaling as $1/N^2$ in definite conditions, and forming the Heisenberg 
limit or superefficiency property \cite{DAriano01,Giovannetti04,Giovannetti06}.
Various schemes with separable preparation or measurement versus entangled preparation or 
measurement are reviewed in 
Refs.~\cite{Giovannetti04,Giovannetti06,Giovannetti11,Hyllus12,Demkowicz14,Toth14}, 
to report the $1/N^2$ superefficiency resulting from entanglement.

There are however different forms of entanglement, and different forms for the parametric 
dependence on $\xi$, and not all behave equally in relation to the possible improvement by the
$1/N^2$ scaling. When the parameter $\xi$ to be estimated characterizes a unitary transformation
$\mathsf{U}_\xi$ such as a phase or frequency, it is known that the estimation can benefit from 
the $1/N^2$ superefficiency by exploiting entanglement \cite{Giovannetti04,Ji08,Giovannetti11}. 
It has however been found that this superefficiency is very fragile when exposed to decoherence 
or quantum noise. Quantum estimation, initiated in noise-free conditions, has more recently been 
investigated in the presence of noise or decoherence 
\cite{Huelga97,Giovannetti11,Escher11,Demkowicz14}. 
It has been realized that for estimating a parameter $\xi$ of a unitary $\mathsf{U}_\xi$, any 
small amount of depolarizing noise is sufficient, for the efficiency, to ruin the $1/N^2$ 
scaling of an entangled process, in the asymptotic limit of large $N$, and return it to the 
$1/N$ scaling of the separable process \cite{Ji08,Giovannetti11,Demkowicz12}.

To complement this asymptotic characterization, in the present paper we characterize how the 
estimation efficiency evolves as a function of the size $N$ of the entangled system and also of 
the degree 
of entanglement. As an important generic estimation task, we consider estimating the phase $\xi$ 
of a unitary transformation $\mathsf{U}_\xi$ acting on a qubit, yet with a characterization which 
is meaningful as well for frequency estimation. Decoherence on the qubit is represented by the 
broad class of noises commuting with the phase rotation, which includes depolarizing, phase-flip, 
and thermal quantum noises. The efficiency of estimation from noisy qubits is assessed by the 
quantum Fisher information $F_q(\xi)$. For this generic estimation task, along the same line,
previous studies examined the case of separable qubits \cite{Chapeau16} or the effect of an 
entangled qubit pair in the presence of depolarizing noise \cite{Chapeau17}.
Now in the present report, estimation takes place on any number $N$ of qubits in the presence of a 
broader class of quantum noises. The quantum Fisher information $F_q(\xi)$ is derived for any 
system size $N$ and degree of entanglement; $F_q(\xi)$ can then be analyzed in the various noise 
configurations and, in particular, it allows us to identify the existence of optimal size and 
degree of entanglement maximizing the estimation efficiency.

\section{The estimation task}

A quantum system in a $D$-dimensional Hilbert space $\mathcal{H}_D$ has its state represented 
by the density operator $\rho_\xi$ dependent upon an unknown parameter $\xi$. When measuring 
$\rho_\xi$ for estimating $\xi$, the overall best efficiency is controlled by the quantum 
Fisher information $F_q(\xi )$ contained in the density operator $\rho_\xi$ about the parameter 
$\xi$ \cite{Braunstein94,Paris09}. By referring to the eigendecomposition of $\rho_\xi$ in its 
orthonormal eigenbasis $\rho_\xi=\sum_{j=1}^D \lambda_j \ket{\lambda_j}\bra{\lambda_j}$, 
one has access to the expression \cite{Holevo82,Paris09,Chapeau15}
\begin{equation}
F_q(\xi ) = 2\sum_{j,k} 
\dfrac{|\braket{\lambda_j |\partial_\xi \rho_\xi| \lambda_k}|^2}{\lambda_j + \lambda_k} \;,
\label{Fq1}
\end{equation}
where the sums in Eq.~(\ref{Fq1}) include all terms corresponding to eigenvalues 
$\lambda_j + \lambda_k \not = 0$. 

For the qubit with two-dimensional Hilbert space $\mathcal{H}_2$, which is a fundamental system 
of quantum information, we consider the essential task of estimating the phase $\xi$ of a generic 
transformation defined by the unitary operator 
\begin{equation}
\mathsf{U}_\xi=\exp\Bigl(-i\dfrac{\xi}{2} \vec{n} \cdot \vec{\sigma} \Bigr) \;,
\label{Uxi1}
\end{equation}
where $\vec{n}$ is a real unit vector of $\mathbbm{R}^3$ which is given,
and $\vec{\sigma}=[\sigma_x, \sigma_y, \sigma_z]$ is a formal vector assembling the three 
(unitary Hermitian) Pauli operators $\sigma_x =\ket{0}\bra{1}+\ket{1}\bra{0}$, 
$\sigma_y =i\ket{1}\bra{0}-i\ket{0}\bra{1}$ and $\sigma_z =\ket{0}\bra{0}-\ket{1}\bra{1}$.
For the estimation, an initial qubit acting as a probe 
is prepared in the quantum state with density operator $\rho_0$ experiencing the transformation 
$\rho_0 \mapsto \mathsf{U}_\xi \rho_0 \mathsf{U}_\xi^\dagger$. For more realistic conditions, 
we consider that the $\xi$-dependent transformed state 
$\mathsf{U}_\xi \rho_0 \mathsf{U}_\xi^\dagger$ is not directly accessible to measurement for 
estimating $\xi$, but after the action of a quantum noise represented by the completely positive
trace-preserving linear superoperator $\mathcal{N}(\cdot)$. The $\xi$-dependent noisy state 
$\rho_\xi=\mathcal{N}(\mathsf{U}_\xi \rho_0 \mathsf{U}_\xi^\dagger)$, when used for estimating
$\xi$, is then associated with the overall best efficiency controlled by the quantum Fisher 
information $F_q(\xi )$ of Eq.~(\ref{Fq1}).

For the qubit, it is convenient to refer to the Bloch representation \cite{Nielsen00,Wilde13},
where the initial state $\rho_0$ of the probe is characterized by the Bloch vector $\vec{r}_0$ in 
$\mathbbm{R}^3$. Then the transformation by $\mathsf{U}_\xi$ of Eq.~(\ref{Uxi1}) amounts to a 
rotation of $\vec{r}_0$ around the axis $\vec{n}$ by the angle $\xi$ in $\mathbbm{R}^3$. The axis 
$\vec{n}$ can be assigned the coelevation angle $\theta_n$ and azimuth $\varphi_n$ 
characterizing the unit vector 
$\vec{n}=[\sin(\theta_n)\cos(\varphi_n), \sin(\theta_n)\sin(\varphi_n), \cos(\theta_n)]^\top$
of $\mathbbm{R}^3$. We then define in $\mathcal{H}_2$ the pure quantum state  
$\ket{\vec{n}}=\ket{0'}=\cos(\theta_n /2)\ket{0}+e^{i\varphi_n}\sin(\theta_n /2)\ket{1}$  as the 
state having $\vec{n}$ as its Bloch vector. The pure state with Bloch vector $-\vec{n}$ is 
denoted $\ket{-\vec{n}}=\ket{1'}=\sin(\theta_n /2)\ket{0}-e^{i\varphi_n}\cos(\theta_n /2)\ket{1}$ 
and is orthogonal to $\ket{0'}$ in $\mathcal{H}_2$. Then in $\mathcal{H}_2$ the pure state 
$\ket{+'}=(\ket{0'}+\ket{1'})/\sqrt{2}$ has a Bloch vector $\vec{n}_\perp$ in $\mathbbm{R}^3$ 
orthogonal to the rotation axis $\vec{n}$. In Bloch representation, the action of the noise 
$\mathcal{N}(\cdot)$ on the qubit is equivalent to an affine transformation of its Bloch vector 
\cite{Nielsen00} as 
\begin{equation}
\vec{r} \longmapsto A \vec{r} + \vec{c} \;,
\label{Aaffine}
\end{equation}
with $A$ a $3\times 3$ real matrix and $\vec{c}$ a real vector of $\mathbbm{R}^3$, mapping the
Bloch ball onto itself.

To specify the noise acting on the qubit, we choose a noise process which commutes with the 
rotation $\mathsf{U}_\xi$ around the axis $\vec{n}$.
A similar assumption is performed in Ref.~\cite{Shaji07} for rotations around the $Oz$ axis,
``to separate cleanly the effect of the parameter from the effects of decoherence'', when the 
process under estimation, itself, introduces a coherent rotation around this invariant axis.
Especially, this provides the faculty to represent with one single noise operation, separate 
decohering actions which could take place before and after rotation of the probe by 
$\mathsf{U}_\xi$ of Eq.~(\ref{Uxi1}), and which can then be lumped into a single one. 
Commutation with the coherent rotation of the probe by $\mathsf{U}_\xi$, is a 
necessary and sufficient condition for the noise to be lumpable in this way.
We can now transpose the noise model of Ref.~\cite{Shaji07} invariant to rotations around the 
$Oz$ axis, to a noise model with invariance relative to the arbitrary axis $\vec{n}$ here.
Any noise model on the qubit can be completely defined by specifying the action of the 
superoperator $\mathcal{N}(\cdot)$ on a basis for operators on $\mathcal{H}_2$. 
We define the alternative Pauli operators as $\sigma'_x =\ket{0'}\bra{1'}+\ket{1'}\bra{0'}$, 
$\sigma'_y =i\ket{1'}\bra{0'}-i\ket{0'}\bra{1'}$ and 
$\sigma'_z =\ket{0'}\bra{0'}-\ket{1'}\bra{1'}$. Associated with $\mathrm{I}_2$ the identity on 
$\mathcal{H}_2$, the set $\{\mathrm{I}_2, \sigma'_x, \sigma'_y, \sigma'_z \}$ as well as
$\{\mathrm{I}_2, \sigma_x, \sigma_y, \sigma_z \}$ form two orthogonal bases for operators on
$\mathcal{H}_2$. Transposing Eq.~(4.3) of Ref.~\cite{Shaji07}, the quantum noise we consider is 
defined by
\begin{eqnarray}
\label{Ns00}
\mathcal{N} \bigl( \mathrm{I}_2 \bigr) &=& \mathrm{I}_2 + \mu_0 \sigma'_z \;, \\
\label{Ns11}
\mathcal{N} \bigl( \sigma'_z \bigr) &=& \mu_1 \sigma'_z \;, \\
\label{Ns01}
\mathcal{N} \bigl( \sigma'_x \pm i \sigma'_y \bigr) &=& 
\mu_2 e^{\mp i\omega t}  \bigl( \sigma'_x \pm i \sigma'_y \bigr) \;.
\end{eqnarray}
Based on Ref.~\cite{Shaji07}, the noise parameters here are $\mu_1=e^{-t/T_1} \in [0, 1]$ and 
$\mu_2=e^{-t/T_2} \in [0, 1]$, with a longitudinal relaxation time $T_1$ and a transverse 
dephasing time $T_2$ satisfying $T_2 \le 2T_1$ for complete positivity, equivalent to
$\mu_2^2 \le \mu_1$; also $\mu_0=\mu(1-\mu_1)$ with $\mu\in [-1, 1]$. Following 
Ref.~\cite{Shaji07}, the noise model of Eqs.~(\ref{Ns00})--(\ref{Ns01}) represents the most 
general noise model independently acting on single qubits, with here an arbitrary invariance axis 
$\vec{n}$, and an underlying temporal dynamics stationary and continuously differentiable.
With this generality, the noise model incorporates in Eq.~(\ref{Ns01}) an angle $\omega t$ 
representing the possibility of a coherent rotation around $\vec{n}$ contributed by the noise.
Such a coherent rotation, if present, leads any procedure for estimating the phase shift $\xi$ 
accompanied with a noise interaction over the duration $t$, to actually estimate $\omega t + \xi$.
As such, this coherent rotation by $\omega t$ has no influence on the statistical fluctuation 
limiting the estimation accuracy; accordingly, as we shall see, it has no impact on the quantum 
Fisher information $F_q(\xi)$ quantifying the estimation efficiency.

When the Bloch vectors in $\mathbbm{R}^3$ are referred to the orthonormal basis 
$\{\vec{n}_\perp, \vec{n}'_\perp = \vec{n} \times \vec{n}_\perp, \vec{n} \}$, the quantum
noise of Eqs.~(\ref{Ns00})--(\ref{Ns01}) has in Eq.~(\ref{Aaffine}) the matrix
\begin{equation}
A =
\left[ \begin{array}{ccc} 
\mu_2\cos(\omega t) & -\mu_2\sin(\omega t) & 0 \\
\mu_2\sin(\omega t) &  \mu_2\cos(\omega t) & 0 \\
0                   &  0                   & \mu_1 
\end{array} \right] \;,
\label{Aaffine2}
\end{equation}
and the vector $\vec{c}=[0, 0, \mu_0]^\top$. From the form of $A$ in Eq.~(\ref{Aaffine2}), the 
action of the noise is to compress the Bloch vector by $\mu_1$ in the direction $\vec{n}$ and by 
$\mu_2$ in the plane $(\vec{n}_\perp, \vec{n}'_\perp)$ orthogonal to $\vec{n}$, with larger 
compression as a larger level of noise. At very long exposition time $t \rightarrow \infty$ to the 
noise, $\mu_1$ and $\mu_2$ go to zero effecting maximum compression, while the noisy qubit relaxes 
to the equilibrium state with Bloch vector $\vec{c}=\mu_0 \vec{n}$.
From $A$ and $\vec{c}$, one has access to an explicit expression for the quantum Fisher 
information $F_q(\xi)$ of Eq.~(\ref{Fq1}), by means of Eq.~(35) of Ref.~\cite{Chapeau16}, from
which it follows here $F_q(\xi)=(\mu_2 r_{0\perp})^2$, with $r_{0\perp}$ the magnitude of the
component orthogonal to the rotation axis $\vec{n}$ of the Bloch vector $\vec{r}_0$ 
characterizing the initial state $\rho_0$ of the probe.
Consequently, with the noise model of Eqs.~(\ref{Ns00})--(\ref{Ns01}), maximization of the
quantum Fisher information $F_q(\xi)$ is achieved by any probe initialized in a pure
state with a unit Bloch vector $\vec{r}_0$ orthogonal to the rotation axis $\vec{n}$, i.e.\
verifying $(r_{0\perp})^2=\|\vec{r}_0\|^2=1$, to reach the maximum $F_q^{\rm max}(\xi)=\mu_2^2$.
The pure state $\ket{+'}=(\ket{0'}+\ket{1'})/\sqrt{2}$ realizes such an optimal probe, having
the density operator $\rho_0^{\rm opt}=\ket{+'}\bra{+'}$ and Bloch vector 
$\vec{r}_0=\vec{n}_\perp$.
More complicated quantum noises exist, with an optimal probe maximizing $F_q(\xi)$ which is not 
orthogonal to the rotation axis $\vec{n}$ in Bloch representation, as exemplified in 
Ref.~\cite{Chapeau15}. These are noises that do not share the commutation property with the rotations 
around $\vec{n}$ as it holds with the noise model of Eqs.~(\ref{Ns00})--(\ref{Ns01}).

The experiment where the probe qubit is prepared in the optimal state 
$\rho_0^{\rm opt}=\ket{+'}\bra{+'}$ can be 
repeated $N$ times on $N$ independent and identical preparations of the probe qubit. This is 
equivalent to considering a composite probe of $N$ qubits prepared in the separable state 
$(\rho_0^{\rm opt})^{\otimes N}$. In such circumstance the quantum Fisher information is additive
\cite{Ji08}, and achieves the maximum $N \mu_2^2$ associated with $1/N$ scaling of the 
mean-squared estimation error. As announced in the Introduction, better estimation efficiency 
may be accessible by exploiting instead an $N$-qubit probe prepared in an entangled state, as
we now examine.

\section{With $N$ entangled qubits}

To select an efficient yet tractable family of $N$-qubit entangled states, we turn to the Hermitian
operator $\vec{n} \cdot \vec{\sigma}$ appearing in Eq.~(\ref{Uxi1}), which represents a spin
observable with eigenvalues $\pm 1$ and eigenstates $\ket{\pm \vec{n}}$ in $\mathcal{H}_2$. As a 
result, the unitary $\mathsf{U}_\xi$ of Eq.~(\ref{Uxi1}) has the two eigenvalues $\exp(-i \xi/2)$ 
and $\exp(i \xi/2)$ respectively associated with the same two eigenstates $\ket{\vec{n}}=\ket{0'}$ 
and $\ket{-\vec{n}}=\ket{1'}$ in $\mathcal{H}_2$. We then choose for the $N$-qubit input probe in 
$\mathcal{H}_2^{\otimes N}$ the density operator $\rho_0=\ket{\psi_0}\bra{\psi_0}$ with the pure 
quantum state
\begin{equation}
\ket{\psi_0}= \sqrt{1-\kappa} \ket{0'}^{\otimes N} + \sqrt{\kappa}\ket{1'}^{\otimes N}
=\sqrt{1-\kappa} \ket{0'_N} + \sqrt{\kappa}\ket{1'_N} \;,
\label{psi0}
\end{equation}
where $\kappa \in [0, 1]$ is a Schmidt coefficient allowing to control the degree of 
entanglement in $\ket{\psi_0}$, spanning from no entanglement at $\kappa \in \{0, 1 \}$, to
maximal entanglement at $\kappa =1/2$, to partial entanglement at $\kappa \not \in \{0, 1/2, 1 \}$.
It is especially useful to test configurations of partial entanglement, because situations are
known in quantum information where optimal or efficient processing occurs at partial entanglement, 
away from both separable and maximally entangled conditions 
\cite{Wineland92,Huelga97,Ulam01,Chapeau17,Kok17}; and this will be observed here, in definite
conditions.
Since the states $\ket{0'}$ and $\ket{1'}$ are eigenstates of $\mathsf{U}_\xi$ in Eq.~(\ref{Uxi1}), 
the $N$-qubit transformed state is $\ket{\psi_1}=\mathsf{U}_\xi^{\otimes N} \ket{\psi_0} = 
\sqrt{1-\kappa}\, e^{-iN\xi/2} \ket{0'_N} + \sqrt{\kappa}\, e^{iN\xi/2}\ket{1'_N}$, with density 
operator $\rho_1=\ket{\psi_1}\bra{\psi_1}$ as
\begin{equation}
\rho_1 (\xi)=(1-\kappa) \ket{0'_N}\bra{0'_N} + \kappa\ket{1'_N}\bra{1'_N} + 
\sqrt{(1-\kappa)\kappa}\, 
\Bigl( e^{iN\xi} \ket{1'_N}\bra{0'_N} + e^{-iN\xi} \ket{0'_N}\bra{1'_N} \Bigr) \;.
\label{ro1}
\end{equation}
And for the quantum state $\rho_1(\xi)$ of Eq.~(\ref{ro1}), one obtains the derivative
\begin{equation}
\partial_\xi \rho_1 (\xi)=i N \sqrt{(1-\kappa)\kappa}\, 
\Bigl( e^{iN\xi} \ket{1'_N}\bra{0'_N} - e^{-iN\xi} \ket{0'_N}\bra{1'_N} \Bigr) \;.
\label{dro1}
\end{equation}

For applying the noise model of Eqs.~(\ref{Ns00})--(\ref{Ns01}) to the $N$-qubit entangled state of 
Eq.~(\ref{ro1}), it is convenient to use the equivalent characterization by the four 
transformations
\begin{eqnarray}
\label{Nb00}
\mathcal{N} \bigl( \ket{0'}\bra{0'} \bigr) &=& 
\dfrac{1}{2}(1+\alpha_0) \ket{0'}\bra{0'} + \dfrac{1}{2}(1-\alpha_0) \ket{1'}\bra{1'} \;, \\
\label{Nb11}
\mathcal{N} \bigl( \ket{1'}\bra{1'} \bigr) &=& 
\dfrac{1}{2}(1-\alpha_1) \ket{0'}\bra{0'} + \dfrac{1}{2}(1+\alpha_1) \ket{1'}\bra{1'} \;, \\
\label{Nb01}
\mathcal{N} \bigl( \ket{0'}\bra{1'} \bigr) &=& \alpha_2 \ket{0'}\bra{1'} \;, \\
\label{Nb10}
\mathcal{N} \bigl( \ket{1'}\bra{0'} \bigr) &=& \alpha_2^* \ket{1'}\bra{0'} \;,
\end{eqnarray}
with the parameters $\alpha_0 =\mu_1 +\mu_0$ and $\alpha_1 =\mu_1 -\mu_0$, and also
$\alpha_2 =\mu_2 e^{-i\omega t}$.

From the one-qubit characterization of Eqs.~(\ref{Nb00})--(\ref{Nb10}), when the noise acts 
independently on each qubit, its action extends to $N$-qubit states by $N$-fold tensor product. 
For instance, for the $N$-qubit state $\ket{0'_N}\bra{1'_N}$ appearing in $\rho_1(\xi)$ of 
Eq.~(\ref{ro1}), one has $\ket{0'_N}\bra{1'_N}=(\ket{0'}\bra{1'})^{\otimes N}$, and from 
Eq.~(\ref{Nb01}) one can write
$\mathcal{N}^{\otimes N} \bigl( \ket{0'_N}\bra{1'_N} \bigr)=
\mathcal{N}^{\otimes N} \bigl( (\ket{0'}\bra{1'})^{\otimes N} \bigr)=
\bigl( \mathcal{N}(\ket{0'}\bra{1'}) \bigr)^{\otimes N}=
\alpha_2^N (\ket{0'}\bra{1'})^{\otimes N} = \alpha_2^N \ket{0'_N}\bra{1'_N} $.

In this way, the action of the quantum noise on the $N$-qubit state $\rho_1(\xi)$ of 
Eq.~(\ref{ro1}), produces the $N$-qubit noisy state $\rho_\xi=\mathcal{N}^{\otimes N}(\rho_1)$ 
which follows from Eqs.~(\ref{Nb00})--(\ref{Nb10}). By linearity of the superoperator
$\mathcal{N}(\cdot)$, we have 
for the derivative $\partial_\xi \rho_\xi = \partial_\xi\mathcal{N}^{\otimes N}(\rho_1) 
=\mathcal{N}^{\otimes N}(\partial_\xi \rho_1)$ which applied to Eq.~(\ref{dro1}) gives
\begin{equation}
\partial_\xi \rho_\xi=i N \sqrt{(1-\kappa)\kappa}\; 
\Bigl( \alpha_2^{*N} e^{iN\xi} \ket{1'_N}\bra{0'_N} - \alpha_2^{N} e^{-iN\xi} \ket{0'_N}\bra{1'_N} 
\Bigr) \;.
\label{dro_xi}
\end{equation}

To characterize the quantum Fisher information $F_q(\xi)$ of Eq.~(\ref{Fq1}), it is not
enough to characterize the differentiated state $\partial_\xi \rho_\xi$ as accomplished in 
Eq.~(\ref{dro_xi}). One also needs to characterize the $N$-qubit noisy state $\rho_\xi$
with its eigenvalues $\lambda_j$ and eigenstates $\ket{\lambda_j}$, for $j=1$ to $2^N$.

From Eq.~(\ref{ro1}), as a first step, the noisy state $\rho_\xi=\mathcal{N}^{\otimes N}(\rho_1)$ 
can be expressed as
\begin{equation}
\rho_\xi=(1-\kappa) \mathcal{N}^{\otimes N}\bigl( \ket{0'_N}\bra{0'_N} \bigr) + \kappa
\mathcal{N}^{\otimes N}\bigl( \ket{1'_N}\bra{1'_N} \bigr) + \sqrt{(1-\kappa)\kappa}\;
\Bigl( \alpha_2^{*N}
e^{iN\xi} \ket{1'_N}\bra{0'_N} + \alpha_2^{N} e^{-iN\xi} \ket{0'_N}\bra{1'_N} \Bigr) \;.
\label{ro_xi}
\end{equation}
The $N$-qubit noisy state $\mathcal{N}^{\otimes N}\bigl( \ket{0'_N}\bra{0'_N} \bigr)$ is obtained 
with an $N$-fold tensor product of Eq.~(\ref{Nb00}) as 
$\mathcal{N}^{\otimes N}\bigl( \ket{0'_N}\bra{0'_N} \bigr)=
\bigl( \mathcal{N}(\ket{0'}\bra{0'}) \bigr)^{\otimes N}$ yielding
\begin{equation}
\mathcal{N}^{\otimes N}\bigl( \ket{0'_N}\bra{0'_N} \bigr)= \sum_{\vec{u} \in \{0', 1' \}^N} 
\biggl[ \dfrac{1}{2}(1+\alpha_0)\biggr]^{N-w(\vec{u})} 
\biggl[ \dfrac{1}{2}(1-\alpha_0)\biggr]^{w(\vec{u})}
\ket{\vec{u}}\bra{\vec{u}} \;,
\label{N00}
\end{equation}
where the summation extends over all $N$-bit binary sequences $\vec{u}$ made from the bits 
$\{0', 1' \}$ among the $2^N$ possible such sequences, and $w(\vec{u})$ denotes the Hamming 
weight or number between $0$ and $N$ of bits $1'$ in the sequence $\vec{u}$. In a similar way,
the $N$-qubit noisy state $\mathcal{N}^{\otimes N}\bigl( \ket{1'_N}\bra{1'_N} \bigr)$ is obtained 
with an $N$-fold tensor product of Eq.~(\ref{Nb11}) as 
$\mathcal{N}^{\otimes N}\bigl( \ket{1'_N}\bra{1'_N} \bigr)=
\bigl( \mathcal{N}(\ket{1'}\bra{1'}) \bigr)^{\otimes N}$ yielding
\begin{equation}
\mathcal{N}^{\otimes N}\bigl( \ket{1'_N}\bra{1'_N} \bigr)= \sum_{\vec{u} \in \{0', 1' \}^N} 
\biggl[ \dfrac{1}{2}(1-\alpha_1)\biggr]^{N-w(\vec{u})} 
\biggl[ \dfrac{1}{2}(1+\alpha_1)\biggr]^{w(\vec{u})}
\ket{\vec{u}}\bra{\vec{u}} \;.
\label{N11}
\end{equation}

It can be noted from Eqs.~(\ref{N00}) and (\ref{N11}) that 
$\mathcal{N}^{\otimes N}\bigl( \ket{0'_N}\bra{0'_N} \bigr)$ and 
$\mathcal{N}^{\otimes N}\bigl( \ket{1'_N}\bra{1'_N} \bigr)$ contain only diagonal terms under the 
form $\ket{\vec{u}}\bra{\vec{u}}$. In this way, in the $N$-qubit noisy state $\rho_\xi$ of 
Eq.~(\ref{ro_xi}), only the two terms made with $\ket{1'_N}\bra{0'_N}$ and $\ket{0'_N}\bra{1'_N}$ 
are off-diagonal terms of $\rho_\xi$.

Moreover, by orthogonality of the $2^N$ normalized state vectors $\ket{\vec{u}}$, one obtains 
from Eq.~(\ref{ro_xi}) for any particular $\vec{u}$ differing from $0'_N$ and $1'_N$,
\begin{eqnarray}
\nonumber
\rho_\xi \ket{\vec{u}} &=& \Biggl(
(1-\kappa) \biggl[ \dfrac{1}{2}(1+\alpha_0)\biggr]^{N-w(\vec{u})} 
\biggl[ \dfrac{1}{2}(1-\alpha_0)\biggr]^{w(\vec{u})} + \mbox{} \\
\label{ro_u}
&& \quad \kappa
\biggl[ \dfrac{1}{2}(1-\alpha_1)\biggr]^{N-w(\vec{u})} 
\biggl[ \dfrac{1}{2}(1+\alpha_1)\biggr]^{w(\vec{u})} \Biggr) \ket{\vec{u}} \;,
\end{eqnarray}
establishing any such $\ket{\vec{u}}$ differing from $\ket{0'_N}$ and $\ket{1'_N}$ as an 
eigenstate of $\rho_\xi$ with the corresponding eigenvalue readable from Eq.~(\ref{ro_u}).
This determines $2^N-2$ (mutually orthogonal) eigenstates for $\rho_\xi$, which together
span the $(2^N-2)$-dimensional subspace orthogonal to the two-dimensional subspace spanned by 
$\bigl(\ket{0'_N}, \ket{1'_N} \bigr)$. Since $\rho_\xi$ is Hermitian and has therefore a total of 
$2^N$ mutually orthogonal eigenstates, the two remaining eigenstates of $\rho_\xi$ are to be 
found in this two-dimensional subspace spanned by $\bigl(\ket{0'_N}, \ket{1'_N} \bigr)$.
We look for them under the form
\begin{equation}
\ket{\lambda}=a_0 \ket{0'_N} + a_1 \ket{1'_N} \;.
\label{lam1}
\end{equation}
The transformation by $\rho_\xi$ of such a $\ket{\lambda}$ occurs in the plane
$\bigl(\ket{0'_N}, \ket{1'_N} \bigr)$ and can be described by the matrix operation
\begin{equation}
\left[ \begin{array}{cc}
\rho_{00}   & \rho_{01} \\
\rho_{01}^* & \rho_{11}
\end{array} \right]
\left[ \begin{array}{c}
a_0 \\
a_1
\end{array} \right]
= \rho_\xi^{(2)} \ket{\lambda}\;,
\label{Uxi2}
\end{equation}
with the $2\times 2$ matrix $\rho_\xi^{(2)}$ defining the operation of the operator $\rho_\xi$
restricted to the plane $\bigl(\ket{0'_N}, \ket{1'_N} \bigr)$. The form of Eq.~(\ref{Uxi2}) and 
the four matrix elements of $\rho_\xi^{(2)}$ are derived from Eq.~(\ref{ro_xi}) acting on state 
vectors belonging to the plane $\bigl(\ket{0'_N}, \ket{1'_N} \bigr)$ like $\ket{\lambda}$ of 
Eq.~(\ref{lam1}). In this way, one obtains
\begin{eqnarray}
\label{ro_00}
\rho_{00} &=& (1-\kappa) \biggl[ \dfrac{1}{2}(1+\alpha_0)\biggr]^N +
              \kappa     \biggl[ \dfrac{1}{2}(1-\alpha_1)\biggr]^N \;, \\ 
\label{ro_11}
\rho_{11} &=& (1-\kappa) \biggl[ \dfrac{1}{2}(1-\alpha_0)\biggr]^N +
              \kappa     \biggl[ \dfrac{1}{2}(1+\alpha_1)\biggr]^N \;, \\ 
\label{ro_01}
\rho_{01} &=& \sqrt{(1-\kappa)\kappa}\, \alpha_2^N  e^{-iN\xi} \;. 
\end{eqnarray}
We are faced with an eigendecomposition restricted to the plane 
$\bigl(\ket{0'_N}, \ket{1'_N} \bigr)$, with a characteristic equation 
$\det\bigl(\rho_\xi^{(2)} -\lambda \mathrm{I}_2 \bigr)=0$ equivalent to
$\lambda^2- (\rho_{00}+\rho_{11})\lambda + \rho_{00}\rho_{11}-\rho_{01}\rho_{01}^* =0$.
This second-degree equation in $\lambda$ has the discriminant 
$\Delta= (\rho_{00}-\rho_{11})^2 +4|\rho_{01}|^2$, and two roots providing the two (real)
eigenvalues that we seek as $\lambda_\pm = (\rho_{00}+\rho_{11} \pm \sqrt{\Delta} )/2$. 
Especially useful to us in the sequel is the sum $\lambda_+ + \lambda_-=\rho_{00}+\rho_{11}$, 
i.e.\
\begin{equation}
\lambda_+ + \lambda_- = (1-\kappa)\beta_0 + \kappa\beta_1 \;,
\label{lam1_lam2}
\end{equation}
with the two parameters
\begin{eqnarray}
\label{beta1}
\beta_0 &=& \biggl[ \dfrac{1}{2}(1+\alpha_0)\biggr]^N +
            \biggl[ \dfrac{1}{2}(1-\alpha_0)\biggr]^N \;, \\
\label{beta2}
\beta_1 &=& \biggl[ \dfrac{1}{2}(1+\alpha_1)\biggr]^N +
            \biggl[ \dfrac{1}{2}(1-\alpha_1)\biggr]^N \;.
\end{eqnarray}

Now that the two eigenvalues $\lambda_\pm$ are expressed, the two corresponding eigenstates
$\ket{\lambda_\pm}$ readily follow by solving the two linear systems 
$\rho_\xi^{(2)} \ket{\lambda_\pm}= \lambda_\pm \ket{\lambda_\pm}$. We do not write explicitly
the resulting expressions, knowing that for the sequel we shall see that it is enough to use 
the characterization $\ket{\lambda_+}=a_0 \ket{0'_N} + a_1 e^{iN(\omega t+ \xi)} \ket{1'_N}$ and
$\ket{\lambda_-}=a_1e^{-iN(\omega t+ \xi)} \ket{0'_N} - a_0 \ket{1'_N}$, with $a_0$ and $a_1$ real, 
for two orthonormal eigenstates $\ket{\lambda_\pm}$ lying in the plane 
$\bigl(\ket{0'_N}, \ket{1'_N} \bigr)$.

Now for our objective of computing the quantum Fisher information $F_q(\xi)$ of Eq.~(\ref{Fq1}), 
due to the form of $\partial_\xi \rho_\xi$ in Eq.~(\ref{dro_xi}) involving only the two basis 
operators $\ket{0'_N}\bra{1'_N}$ and $\ket{1'_N}\bra{0'_N}$, it is clear that only the two 
eigenstates $\ket{\lambda_\pm}$ lying in the plane $\bigl(\ket{0'_N}, \ket{1'_N} \bigr)$ can 
contribute a non-vanishing scalar $\braket{\lambda_j |\partial_\xi \rho_\xi| \lambda_k}$ in 
Eq.~(\ref{Fq1}).
Moreover, with the above form determined for $\ket{\lambda_\pm}$, it follows that each of the
two rectangular terms $\braket{\lambda_+ |\partial_\xi \rho_\xi| \lambda_+}$ and
$\braket{\lambda_- |\partial_\xi \rho_\xi| \lambda_-}$ is identically zero. Only the two
diagonal terms $\braket{\lambda_+ |\partial_\xi \rho_\xi| \lambda_-}=
\braket{\lambda_- |\partial_\xi \rho_\xi| \lambda_+}^*$ do not vanish, but yield
\begin{equation}
\braket{\lambda_+ |\partial_\xi \rho_\xi| \lambda_-}=i N \sqrt{(1-\kappa)\kappa}\, 
\alpha_2^N e^{-iN\xi}\;.
\label{ladla}
\end{equation}

Relying on the sum $\lambda_+ + \lambda_-$ of Eq.~(\ref{lam1_lam2}), the quantum Fisher 
information $F_q(\xi)$ of Eq.~(\ref{Fq1}) finally evaluates to
\begin{equation}
F_q(\xi) = 
\dfrac{4(1-\kappa)\kappa N^2 \mu_2^{2N}}{(1-\kappa)\beta_0 + \kappa\beta_1 } \;.
\label{Fq2g}
\end{equation}

It is observed that, by exploiting an $N$-qubit probe in the state of Eq.~(\ref{psi0}) matched to 
the rotation axis $\vec{n}$ through the two qubit states $\ket{0'}$ and $\ket{1'}$ respectively 
having Bloch vector $\vec{n}$ and $-\vec{n}$, then we obtain in Eq.~(\ref{Fq2g}) a quantum Fisher 
information $F_q(\xi)$ which is independent of the axis $\vec{n}$. If this were not done, the 
quantum Fisher information $F_q(\xi)$ would in general depend on the situation of the probe in 
relation to the rotation axis $\vec{n}$. This can be seen starting with the case of a one-qubit 
probe, having $F_q(\xi)=(\mu_2 r_{0\perp})^2$ as explained in the paragraph of 
Eq.~(\ref{Aaffine2}), which is a Fisher information that changes with the angle between $\vec{n}$ 
and the Bloch vector $\vec{r}_0$ characterizing the probe.

In addition, the quantum Fisher information $F_q(\xi)$ of Eq.~(\ref{Fq2g}) is also found 
independent of the unknown phase angle $\xi$ we want to estimate for the qubit rotation around 
the axis $\vec{n}$. This parameter-independent performance is an interesting feature, not 
always present for parametric estimation, ensuring that the overall best estimation efficiency 
assessed by $F_q(\xi)$ is independent of the specific value of the phase $\xi$ being estimated. 
This independence here is related to the qubit noise model having symmetry relative to the 
axis $\vec{n}$ according to Eqs.~(\ref{Ns00})--(\ref{Ns01}) so as to commute with the 
rotation $\mathsf{U}_\xi$ around $\vec{n}$. Other noises with no such 
symmetry would in general lead to a quantum Fisher information $F_q(\xi)$ showing dependence on 
the unknown phase angle $\xi$. This is for instance verified in Ref.~\cite{Chapeau15}, with qubit 
noises not having the above symmetry and with a one-qubit probe showing a quantum Fisher 
information $F_q(\xi)$ with dependence on the unknown phase $\xi$.

In this way, with the broad class of qubit noises of Eqs.~(\ref{Ns00})--(\ref{Ns01}), by selecting 
the probe according to Eq.~(\ref{psi0}), we are able to characterize with Eq.~(\ref{Fq2g}) a 
quantum Fisher information $F_q(\xi)$ which is independent of both $\vec{n}$ and $\xi$, and 
therefore identifies an overall best estimation efficiency uniform for any axis $\vec{n}$ and any 
phase $\xi$.

\section{Analysis of the quantum Fisher information} \label{analN_sec}

The decoherence-free situation with the noise model of Eqs.~(\ref{Ns00})--(\ref{Ns01}) is
characterized by $\mu_0=0$, $\mu_1=\mu_2=1$ and $\omega =0$; this translates for
Eqs.~(\ref{Nb00})--(\ref{Nb01}) into $\alpha_0 =\alpha_1 =\alpha_2=1$, and $\beta_0=\beta_1=1$ in
Eqs.~(\ref{beta1})--(\ref{beta2}). The quantum Fisher information in Eq.~(\ref{Fq2g}) follows as
$F_q(\xi)=4(1-\kappa)\kappa N^2$. This corresponds to the $1/N^2$ scaling of the mean-squared 
estimation error accessible at no noise. This superefficiency is obtained thanks to entanglement 
of the $N$ qubits of the probe. Any non-degenerate degree of entanglement, with 
$\kappa \not \in \{0, 1 \}$ in 
Eq.~(\ref{psi0}), is appropriate to give access to superefficiency with $1/N^2$ scaling of the
mean-squared estimation error. Comparatively, the $N$ independent qubits of the optimal separable 
probe $\ket{+'}^{\otimes N}$ would reach a quantum Fisher information $F_q(\xi)=N$ associated with 
the (less efficient) $1/N$ scaling of the mean-squared estimation error. In this respect, at no 
noise, quantum correlation among the qubits via entanglement is always beneficial to the 
estimation efficiency.

In addition, $F_q(\xi)=4(1-\kappa)\kappa N^2$ is maximized for $\kappa=1/2$ when the prefactor 
$4(1-\kappa)\kappa =1$. This indicates that, at no noise, in the family of probe states of 
Eq.~(\ref{psi0}), it is the maximally entangled state with Schmidt coefficient $\kappa=1/2$ that 
forms the most efficient probe for estimation. We will see that this is not always the case
with noise, where maximum entanglement at $\kappa=1/2$ in Eq.~(\ref{psi0}) is not necessarily
the optimum maximizing $F_q(\xi)$ of Eq.~(\ref{Fq2g}).

\subsection{Large probe size}

In the presence of noise, Eq.~(\ref{Fq2g}) expresses for the quantum Fisher information $F_q(\xi)$ 
a sophisticated dependence with the size $N$ of the entangled probe, that significantly departs 
from simple evolutions as $N^2$ or as $N$. The quantum Fisher information $F_q(\xi)$ of 
Eq.~(\ref{Fq2g}), in accordance with its general definition, is always non-negative. A significant 
property conveyed by Eq.~(\ref{Fq2g}) is that, in the presence of any non-vanishing noise 
according to the model of Eqs.~(\ref{Ns00})--(\ref{Ns01}), the Fisher information $F_q(\xi)$ of 
Eq.~(\ref{Fq2g}) always returns to zero asymptotically at large size $N$ of the entangled probe. 
This is established as follows, when $\mu_0 \ge 0$, given that when $\mu_0 \le 0$ one arrives at
the same conclusion by symmetry of the situation.
When $\mu_0 \ge 0$, one has $0\le \alpha_1 \le \alpha_0 <1$, so that at large $N$, one obtains
for $F_q(\xi)$ of Eq.~(\ref{Fq2g}) the asymptotic form
\begin{equation}
F_q(\xi) \rightarrow 4\kappa 
\dfrac{N^2 \mu_2^{2N}}{[(1+\alpha_0)/2]^N } = 4\kappa
N^2 \left(\dfrac{2\mu_2^2}{1+\mu_1+\mu_0} \right)^N \;.
\label{Fq2g_as}
\end{equation}
The condition $\mu_2^2 \le \mu_1$ guaranteed by complete positivity of the noise, implies that 
$0 \le 2\mu_2^2/(1+\mu_1+\mu_0) <1$ for the exponentiated factor in Eq.~(\ref{Fq2g_as}).
As a consequence, $F_q(\xi)$ controlled by Eq.~(\ref{Fq2g_as}) always decays exponentially to zero 
at large $N \rightarrow \infty$. This means that large sizes $N \rightarrow \infty$ of the 
entangled probe
are always inefficient for estimation, for any configuration of a non-vanishing noise according to 
Eqs.~(\ref{Ns00})--(\ref{Ns01}). Other studies have come to a similar conclusion, with various
types of noise and metrics of efficiency \cite{Ji08,Giovannetti11,Demkowicz12}. We have here
an alternative proof based on the behavior of the quantum Fisher information and holding for the
broad class of noises of Eqs.~(\ref{Ns00})--(\ref{Ns01}).
This outcome materializes the fragility to noise evoked in the 
Introduction for the superefficient estimation. This asymptotic behavior of $F_q(\xi)$ 
indicates that, at large size $N \rightarrow \infty$, the block of $N$ entangled 
qubits behaving in a correlated way gets dominated by the noise and becomes inoperative for 
estimation, and this for any degree of entanglement controlled by the Schmidt coefficient $\kappa$ 
in Eq.~(\ref{psi0}). Comparatively, the optimal separable probe $\ket{+'}^{\otimes N}$ of $N$ 
independent qubits would maintain a quantum Fisher information $F_q(\xi)=N\mu_2^2$ linearly 
increasing with $N$, remaining in this way operative for estimation at any size $N$. Here with 
noise, quantum correlation among the qubits via entanglement becomes detrimental to the estimation 
efficiency for large probe size $N \rightarrow \infty$, at any nonzero level of noise.

This contrasted behavior without or with noise manifests the sophisticated and versatile role of 
quantum correlation by entanglement for information processing. Depending on the conditions, the 
correlated behavior among entangled qubits can reveal beneficial (with no noise) or asymptotically 
detrimental (with noise).

\subsection{Optimal probe size}

Another significant property conveyed by Eq.~(\ref{Fq2g}) is the existence of an optimal value of 
the size $N$ of the entangled probe, able to maximize the Fisher information $F_q(\xi)$ of 
Eq.~(\ref{Fq2g}), this for any configuration of the noise of Eqs.~(\ref{Ns00})--(\ref{Ns01}) and 
any degree of entanglement $\kappa$. An analytical prediction can even be made for an optimal size 
$N_{\rm opt}$ of the entangled probe, by looking for the maximum of the asymptotic form of 
$F_q(\xi)$ given by Eq.~(\ref{Fq2g_as}), which comes out analytically, for any fixed $\kappa$, as
\begin{equation}
N_{\rm opt} =\dfrac{2}{\ln \biggl(\dfrac{1+\mu_1+\mu_0}{2\mu_2^2} \biggr)} \;,
\label{Nopt1}
\end{equation}
this expression being expected as accurate for noise configurations where the optimal
probe size occurs in the range $N_{\rm opt} \gg 1$.

Alternatively, for arbitrary noise configurations, numerical study of $F_q(\xi)$ from 
Eq.~(\ref{Fq2g}), allows one to verify the existence and value of an optimal size $N_{\rm opt}$
of the entangled probe.
This can be exemplified for instance with a noise configuration of great relevance to the
qubit, under the form of the depolarizing noise, which is obtained in 
Eqs.~(\ref{Ns00})--(\ref{Ns01}) when $\mu_0=0$, also $\mu_1=\mu_2=\alpha \in [0, 1]$ and 
$\omega =0$, associated in Eqs.~(\ref{Nb00})--(\ref{Nb01}) with 
$\alpha_0 =\alpha_1 =\alpha_2=\alpha \in [0, 1]$. This describes the situation of a unital (i.e.\ 
verifying $\mathcal{N}( \mathrm{I}_2) = \mathrm{I}_2$) isotropic quantum noise,
with the relaxation times $T_1=T_2$, and thus no privileged axis.

For the depolarizing noise described by the single depolarizing factor $\alpha$, the parameters
$\beta_0$ and $\beta_1$ in Eqs.~(\ref{beta1})--(\ref{beta2}) coincide, so that 
the quantum Fisher information $F_q(\xi)$ of Eq.~(\ref{Fq2g}) reduces to
\begin{equation}
F_q(\xi) = 4(1-\kappa)\kappa \,
\dfrac{N^2 (2\alpha^2)^N}{(1+\alpha)^N + (1-\alpha)^N} \;.
\label{Fq2}
\end{equation}
For the depolarizing noise, Eq.~(\ref{Fq2}) shows that $F_q(\xi)$ is always 
maximized for $\kappa=1/2$ with the prefactor in Eq.~(\ref{Fq2}) culminating at 
$4(1-\kappa)\kappa =1$. This indicates that, in the family of probe states of Eq.~(\ref{psi0}), it 
is always the maximally entangled state at Schmidt coefficient $\kappa=1/2$ that forms the most 
efficient setting for estimation, for any factor $\alpha$ of the depolarizing noise and any 
probe size $N$. 

\begin{figure}[htb]
\centerline{
{\includegraphics[width=56mm]{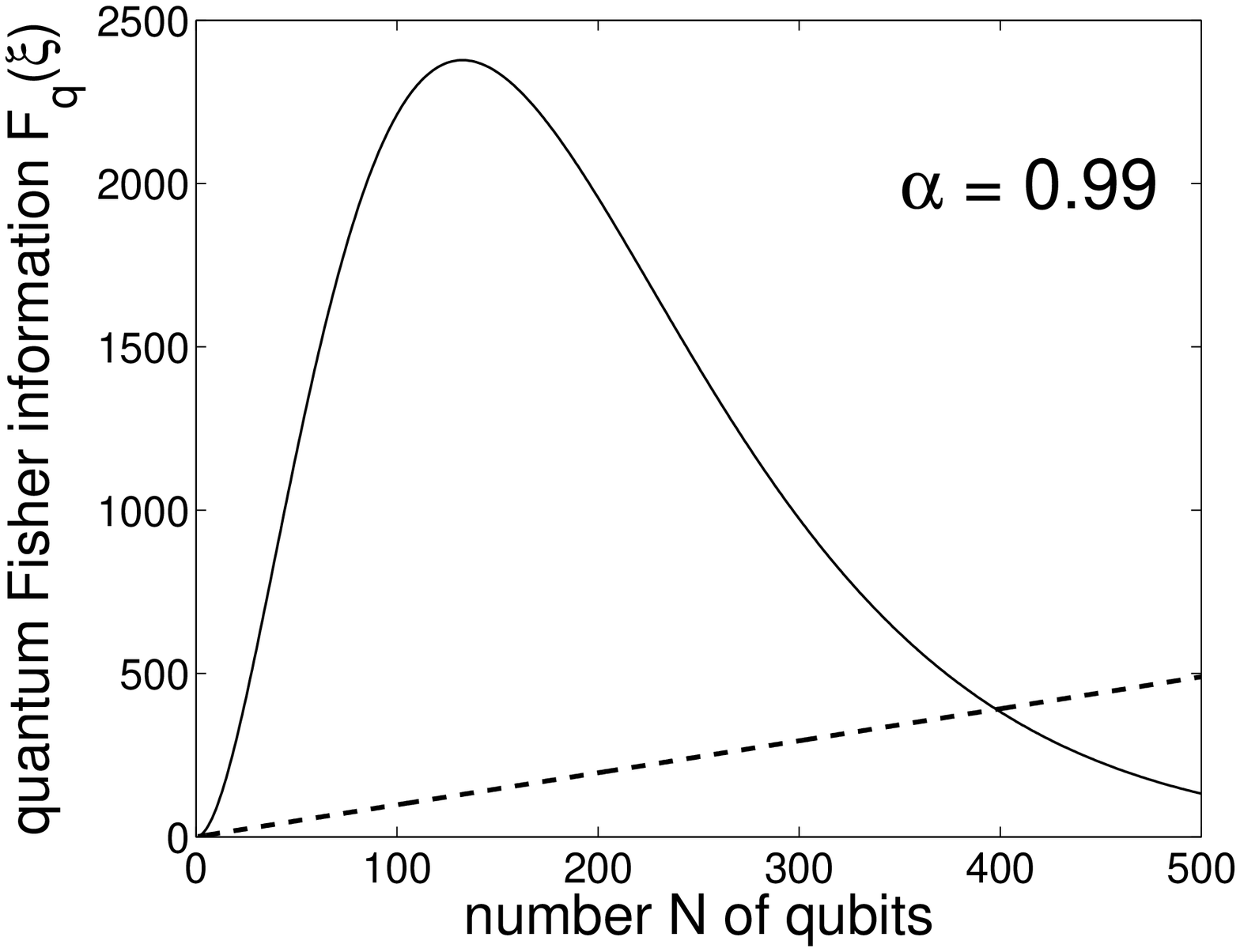}}
{\includegraphics[width=56mm]{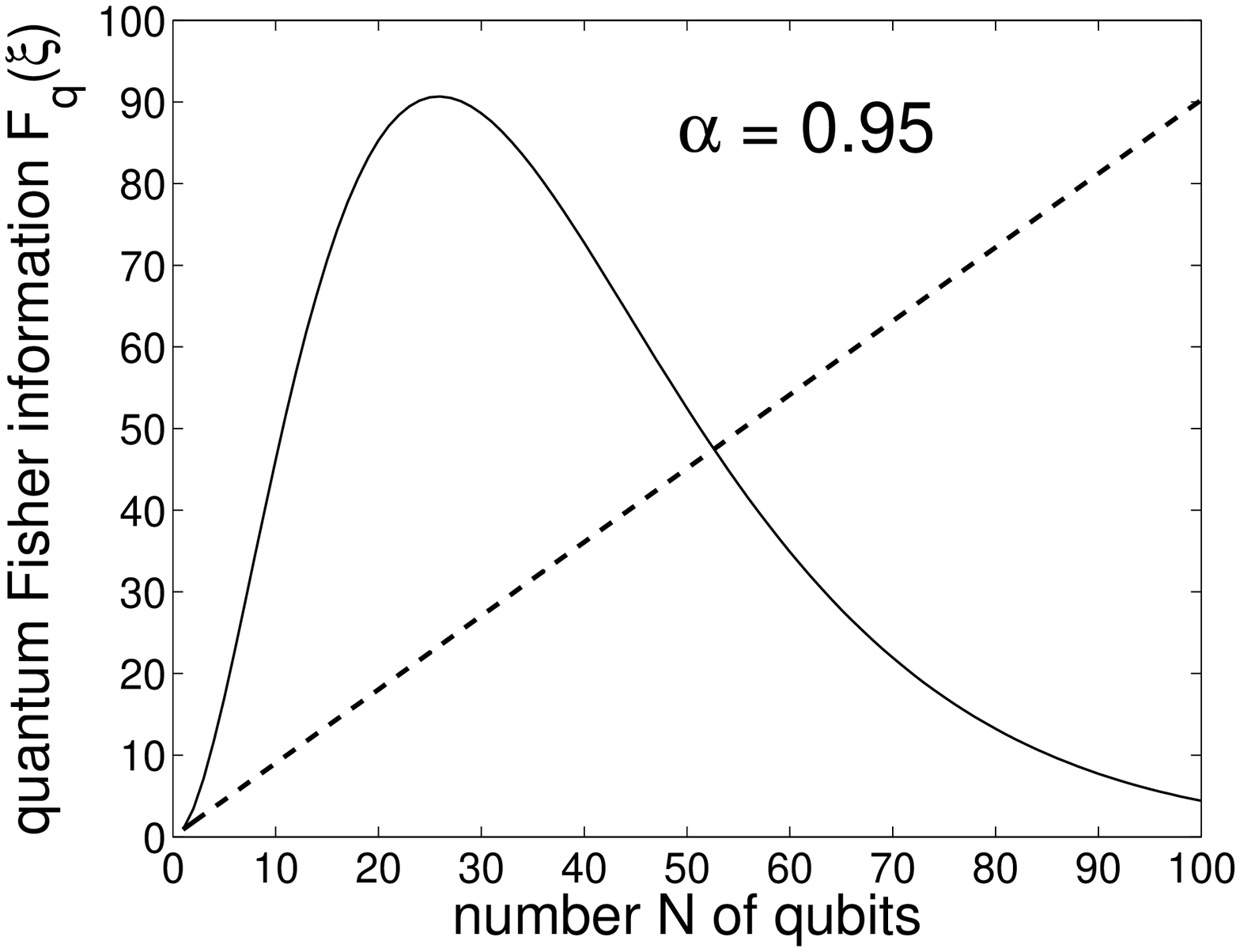}}
{\includegraphics[width=56mm]{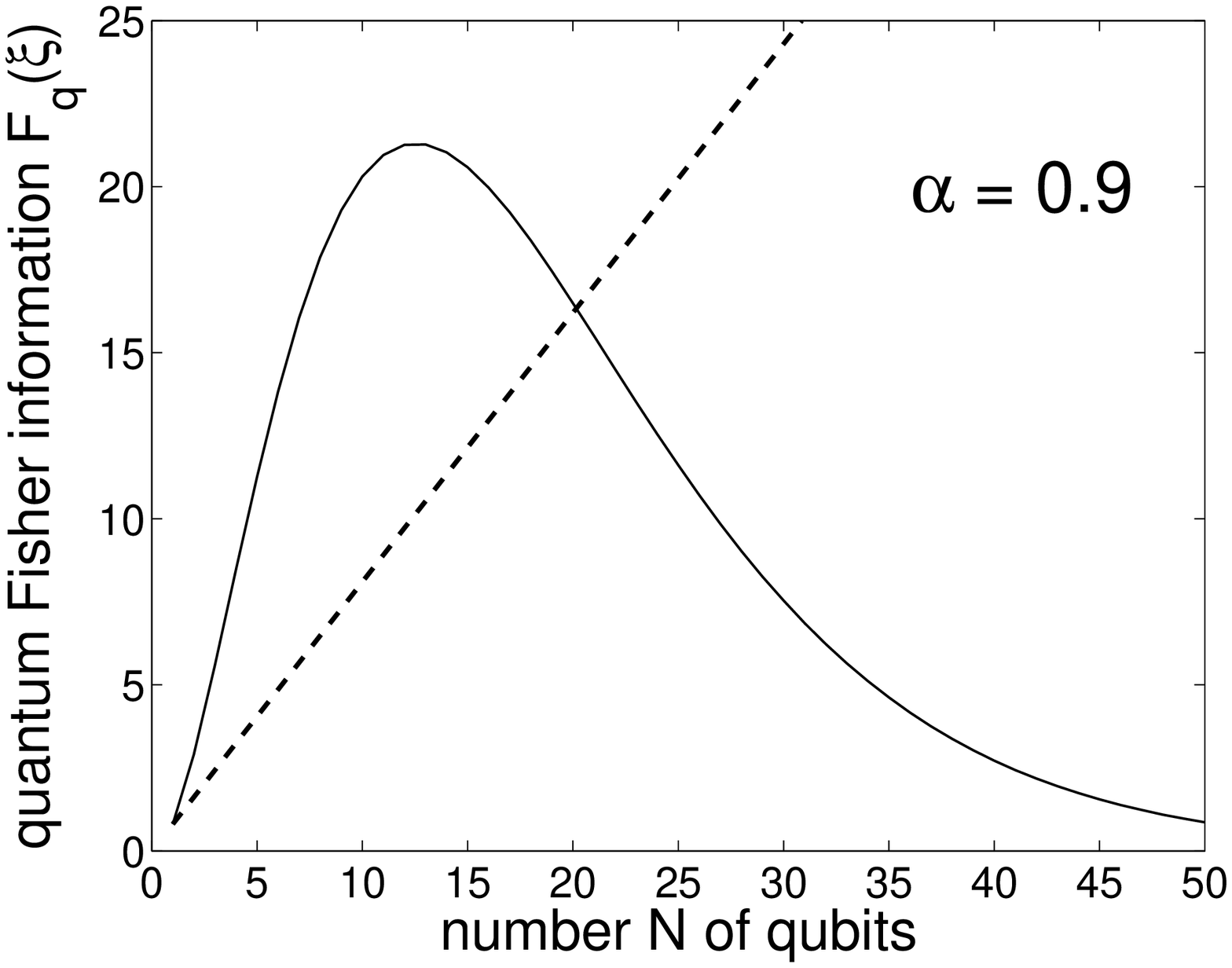}}
}
\caption[what appears in lof LL p177]
{As a function of the number $N$ of qubits in the probe, quantum Fisher information $F_q(\xi)$ 
from Eq.~(\ref{Fq2}) at $\kappa=1/2$ for a maximally entangled probe in Eq.~(\ref{psi0}) 
(solid line), and $F_q(\xi)=N\alpha^2$ for the optimal separable probe $\ket{+'}^{\otimes N}$ 
(dashed line), at $3$ values of the depolarizing noise factor $\alpha$.
}
\label{figFq1}
\end{figure}

For different values of the depolarizing noise factor $\alpha$, 
the quantum Fisher information $F_q(\xi)$ of Eq.~(\ref{Fq2}) at $\kappa=1/2$ is depicted in 
Fig.~\ref{figFq1} as a function of the number $N$ of entangled qubits of the probe.

\begin{figure}[htb]
{\includegraphics[width=84mm]{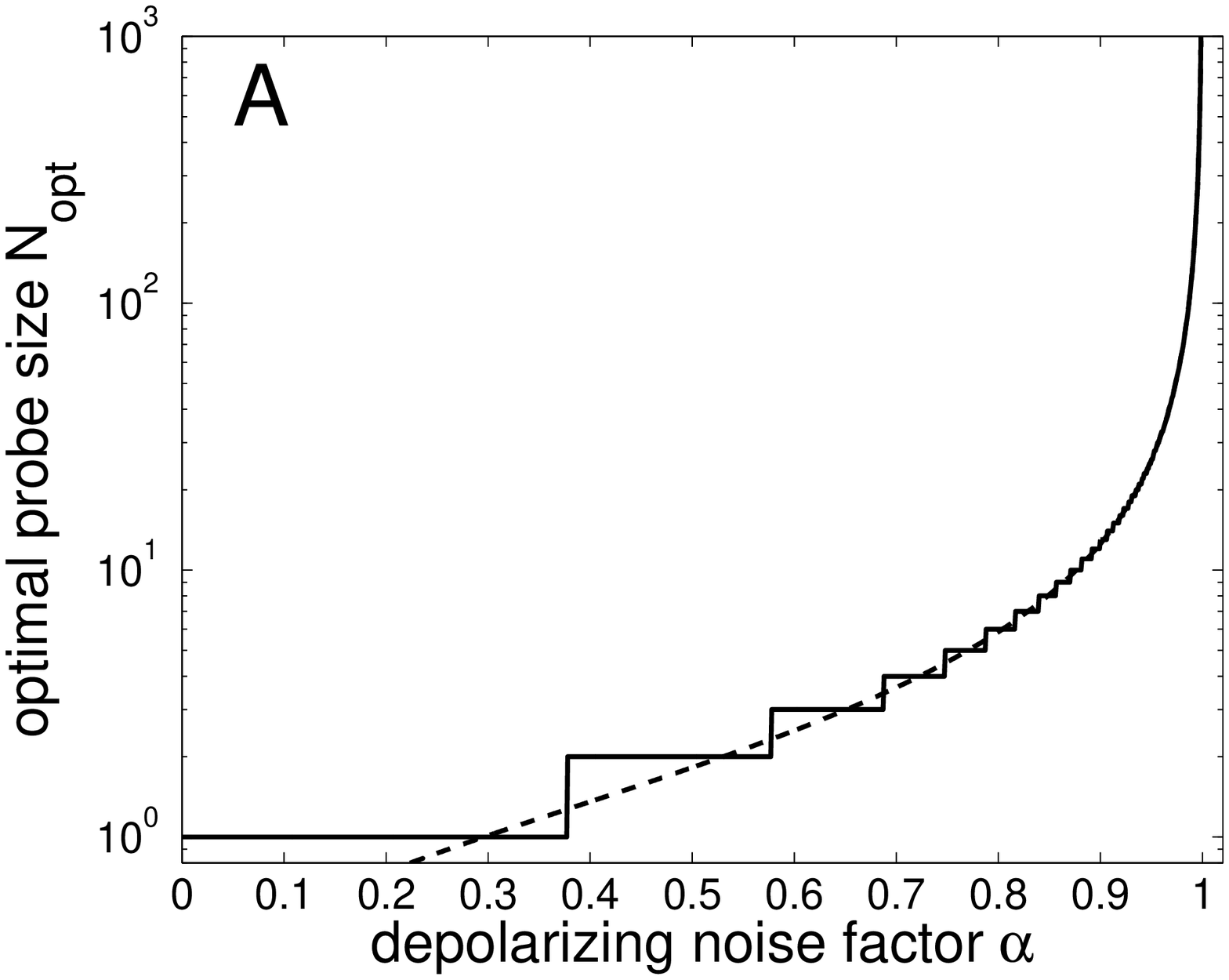}}
\hfill
{\includegraphics[width=84mm]{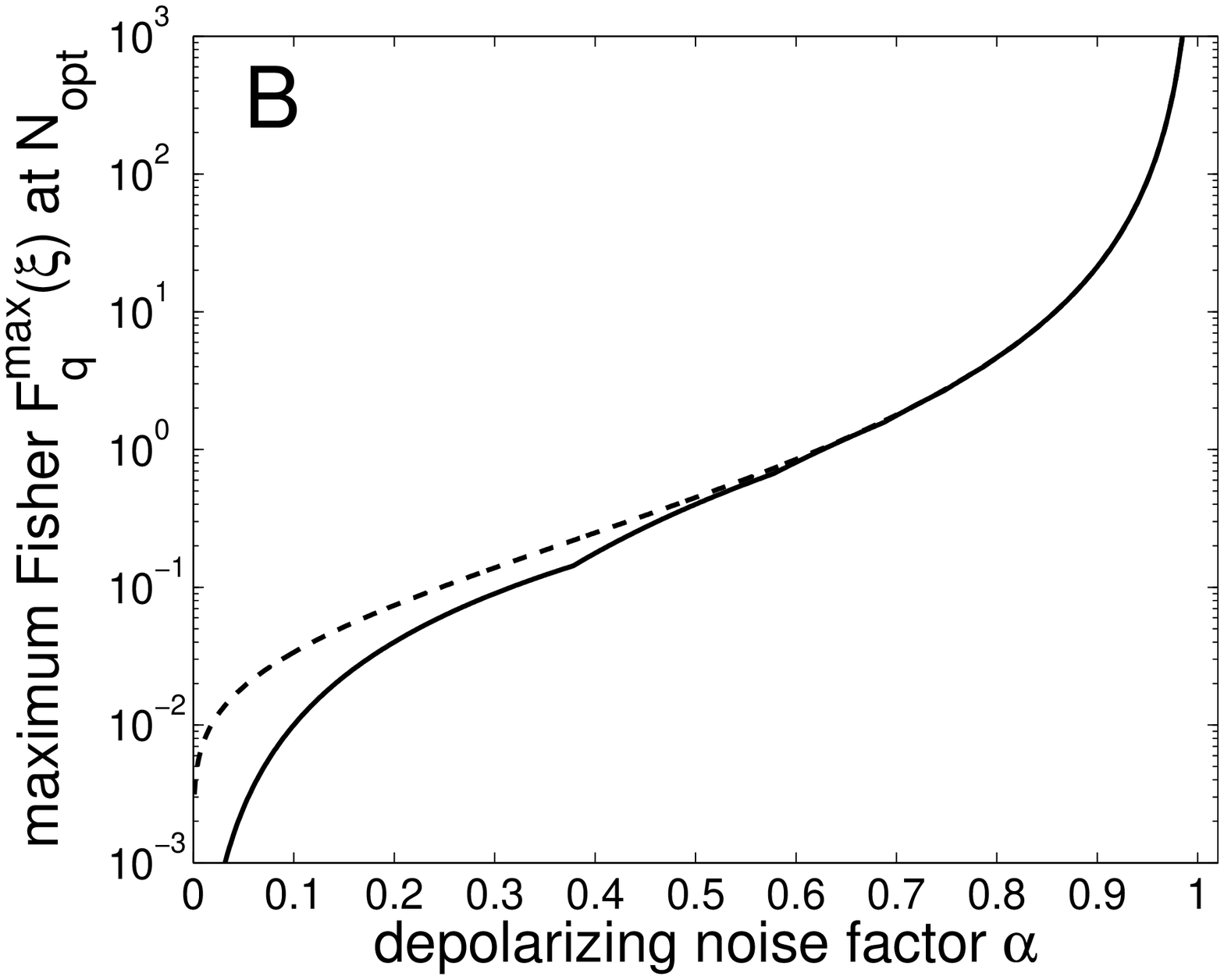}}
\caption[what appears in lof LL p177]
{As a function of the depolarizing noise factor $\alpha$, (A) optimal size $N_{\rm opt}$ of the 
entangled probe maximizing the quantum Fisher information $F_q(\xi)$ of Eq.~(\ref{Fq2})
(solid line) and its analytical expression of Eq.~(\ref{Nopt2}) at small noise when 
$\alpha \rightarrow 1$ (dashed line),
(B) corresponding maximum Fisher information $F_q^{\rm max}(\xi)$ reached in Eq.~(\ref{Fq2}) at 
each $N_{\rm opt}(\alpha)$ when $\kappa=1/2$ (solid line) and its analytical expression from 
Eq.~(\ref{Fq2_as}) at small noise when $\alpha \rightarrow 1$ (dashed line).
}
\label{figNopt}
\end{figure}

Figure~\ref{figFq1} verifies the two generic properties announced for the quantum Fisher 
information $F_q(\xi)$ of Eq.~(\ref{Fq2g}), that $F_q(\xi)$ always returns to zero asymptotically 
at large size $N$ of the entangled probe, and that $F_q(\xi)$ gets maximized at a finite 
(noise-dependent) optimal probe size $N_{\rm opt}$. Figure~\ref{figFq1} also represents the 
Fisher information 
$F_q(\xi)=N\alpha^2$ achieved by $N$ separable qubits in state $\ket{+'}^{\otimes N}$, and clearly 
demonstrates broad ranges of the probe size $N$ where the entangled probe is more efficient than 
the optimal separable probe for estimation, as manifested by a superior Fisher information.
As also visible in Fig.~\ref{figFq1}, the value of the optimal size $N_{\rm opt}$ of the entangled 
probe usually depends on the specific noise configuration.
Figure~\ref{figNopt} represents the optimal size $N_{\rm opt}(\alpha)$ of the entangled probe
maximizing $F_q(\xi)$ of Eq.~(\ref{Fq2}), as a function of the factor $\alpha$ of the 
depolarizing noise, along with the corresponding maximum Fisher information $F_q^{\rm max}(\xi)$ 
reached at $N_{\rm opt}(\alpha)$ in Eq.~(\ref{Fq2}) when $\kappa=1/2$.

Figure~\ref{figNopt}(A) shows an optimal size $N_{\rm opt}(\alpha)$ of the entangled probe which 
increases as the noise factor $\alpha$ increases. At large noise when $\alpha \rightarrow 0$, the
optimal size $N_{\rm opt}$ goes to $1$, expressing in another way that entanglement or correlation
among several qubits ceases to be efficient at large noise. Above $\alpha \approx 0.377$ in 
Fig.~\ref{figNopt}(A) the optimal size $N_{\rm opt}$ of the probe jumps above $1$ and then 
$N_{\rm opt}$ gradually increases towards very large values as $\alpha$ approaches $1$ when the 
noise faints. At small noise when 
$\alpha \rightarrow 1$, entanglement gets more and more efficient, as expressed by larger optimal 
size $N_{\rm opt}$ with increasing $\alpha$. This is also reflected in Fig.~\ref{figNopt}(B) 
showing the maximum Fisher information $F_q^{\rm max}(\xi)$ at $N_{\rm opt}$. This maximum 
$F_q^{\rm max}(\xi)$ also gets very large as $\alpha \rightarrow 1$, expressing larger efficiency 
of entanglement at small noise.

For large size $N$ of the entangled probe with depolarizing noise, Eq.~(\ref{Fq2}) has the 
asymptotic form
\begin{equation}
F_q(\xi) = 4(1-\kappa)\kappa N^2 \left(\dfrac{2\alpha^2}{1+\alpha} \right)^N \;.
\label{Fq2_as}
\end{equation}
In the regime of small noise $\alpha \rightarrow 1$, when the optimal probe size $N_{\rm opt}$ 
is large, the analytical prediction of Eq.~(\ref{Nopt1}) becomes
\begin{equation}
N_{\rm opt}(\alpha) =\dfrac{2}{\ln \biggl(\dfrac{1+\alpha}{2\alpha^2} \biggr)} \;,
\label{Nopt2}
\end{equation}
for maximizing $F_q(\xi)$ of Eq.~(\ref{Fq2_as}), as shown in Fig.~\ref{figNopt}(A).
And when the value $N=N_{\rm opt}$ from Eq.~(\ref{Nopt2}) is placed in Eq.~(\ref{Fq2_as}),
an analytical prediction results for the maximum Fisher information $F_q^{\rm max}(\xi)$ in the 
small-noise regime $\alpha \rightarrow 1$, also shown in Fig.~\ref{figNopt}(B).
As visible in Fig.~\ref{figNopt}, these two analytical predictions from Eqs.~(\ref{Nopt2}) and 
(\ref{Fq2_as}) for the small-noise regime $\alpha \rightarrow 1$, provide a very good match.

\begin{figure}[htb]
\centerline{\includegraphics[width=84mm]{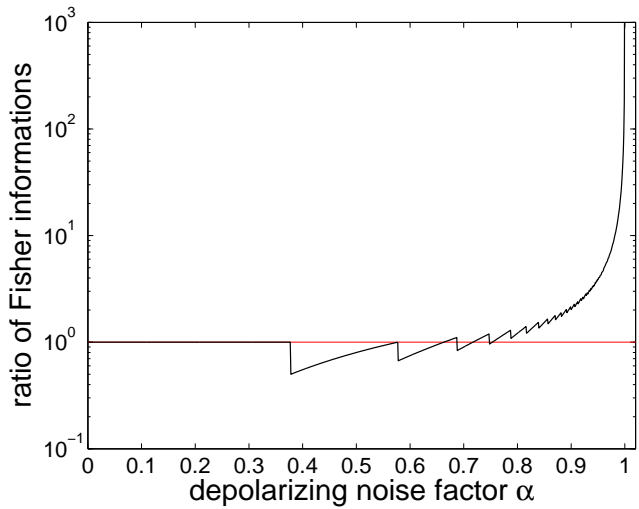}}
\caption[what appears in lof LL p177]
{As a function of the depolarizing noise factor $\alpha$, the ratio 
$F_q^{\rm max}(\xi)/(N_{\rm opt} \alpha^2)$ of the maximum Fisher information 
$F_q^{\rm max}(\xi)$ reached at each $\alpha$ by $N_{\rm opt}(\alpha)$ maximally 
entangled qubits as in Fig.~\ref{figNopt}, to the Fisher information $N_{\rm opt} \alpha^2$ 
reached by the optimal separable probe $\ket{+'}^{\otimes N_{\rm opt}}$ of $N_{\rm opt}$ 
independent qubits. The baseline at ordinate $1$ is shown in red.
}
\label{figFratio}
\end{figure}

The maximum Fisher information $F_q^{\rm max}(\xi)$ at $N_{\rm opt}$ in Fig.~\ref{figNopt} is 
achieved by the entangled probe of Eq.~(\ref{psi0}) at $\kappa=1/2$, and it is meaningful to 
compare it with the Fisher information $N_{\rm opt}\alpha^2$ achieved by the optimal separable 
probe $\ket{+'}^{\otimes N}$ with same size $N=N_{\rm opt}$. This is accomplished in 
Fig.~\ref{figFratio} showing the ratio $F_q^{\rm max}(\xi)/(N_{\rm opt}\alpha^2)$ as a function of 
the depolarizing noise factor $\alpha$.

At large noise when $\alpha \rightarrow 0$ in Fig.~\ref{figFratio} the ratio 
$F_q^{\rm max}(\xi)/(N_{\rm opt}\alpha^2)$ goes to $1$, which indicates a similar efficiency for 
the maximally entangled probe and the separable probe, when they coincide at $N_{\rm opt}=1$ in 
this range of noise. For intermediate $\alpha$ in Fig.~\ref{figFratio}, the ratio can get below $1$, 
indicating some intermediate ranges of noise where the separable probe is in this respect more 
efficient than the entangled probe, at sizes $N_{\rm opt} >1$. However, above 
$\alpha \approx 0.754$ in Fig.~\ref{figFratio}, the ratio always exceeds $1$ and can take very 
large values as the noise vanishes with $\alpha \rightarrow 1$ and the size $N_{\rm opt}$ grows. 
This expresses that at small noise the entangled probe is always more efficient than the separable 
probe, with a superiority which can become arbitrary large as $\alpha \rightarrow 1$ and increasing
size $N_{\rm opt}$ of the entangled probe.

\begin{figure}[htb]
\centerline{\includegraphics[width=84mm]{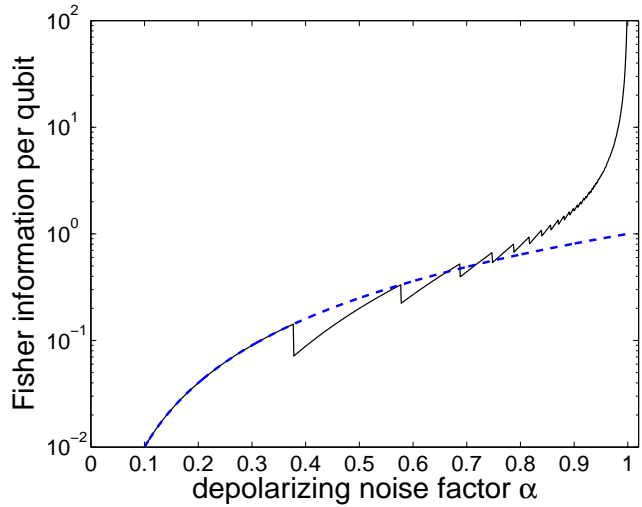}}
\caption[what appears in lof LL p177]
{As a function of the depolarizing noise factor $\alpha$, the maximum Fisher information per 
qubit $F_q^{\rm max}(\xi)/N_{\rm opt}(\alpha)$ for the maximally entangled probe of $N_{\rm opt}$ 
qubits (solid line), and $\alpha^2$ representing the Fisher information per qubit of the optimal 
separable probe $\ket{+'}^{\otimes N_{\rm opt}}$ of $N_{\rm opt}$ independent qubits (dashed 
line).
}
\label{figFperqb}
\end{figure}

A complementary viewpoint is provided by the Fisher information per qubit presented in 
Fig.~\ref{figFperqb}, which amounts to $F_q^{\rm max}(\xi)/N_{\rm opt}(\alpha)$
for the entangled probe and to $\alpha^2$ for the separable probe.

Consistently in Fig.~\ref{figFperqb}, for a noise factor above $\alpha \approx 0.754$ the Fisher 
information per qubit is always superior for the entangled probe compared to the separable probe. 
For the separable probe, the Fisher information per qubit $\alpha^2$ can never exceed $1$. 
By contrast, 
for the entangled probe the Fisher information per qubit $F_q^{\rm max}(\xi)/N_{\rm opt}(\alpha)$ 
can take very large values as $\alpha \rightarrow 1$ when the noise vanishes. This is a striking 
manifestation of the superiority that entanglement can afford over separable states in definite 
circumstances. Here entanglement can bring an arbitrarily large superiority compared to separable 
states at their optimum.

\begin{figure}[htb]
\centerline{\includegraphics[width=84mm]{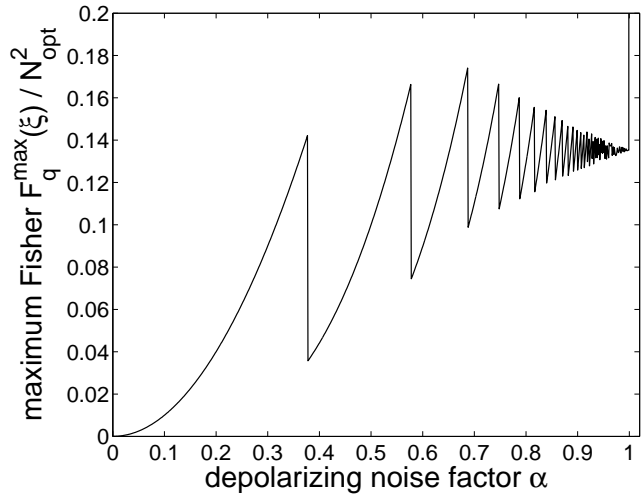}}
\caption[what appears in lof LL p177]
{As a function of the depolarizing noise factor $\alpha$, the ratio
$F_q^{\rm max}(\xi)/N_{\rm opt}^2(\alpha)$ for the maximum Fisher information 
$F_q^{\rm max}(\xi)$ of the maximally entangled probe with $N_{\rm opt}$ qubits,
illustrating the asymptotic trend $F_q^{\rm max}(\xi) \sim 0.135 N_{\rm opt}^2$ as
$\alpha \rightarrow 1$.
}
\label{figFmaxN2}
\end{figure}

In addition, Fig.~\ref{figFmaxN2} represents the ratio $F_q^{\rm max}(\xi)/N_{\rm opt}^2(\alpha)$ 
and shows that this ratio tends to a constant $\approx 0.135$ as the noise diminishes when 
$\alpha \rightarrow 1$.
The analytical expressions provided by Eqs.~(\ref{Nopt2}) and (\ref{Fq2_as}) 
for the small-noise regime $\alpha \rightarrow 1$, lead here to an analytical ratio
$F_q^{\rm max}(\xi)/N_{\rm opt}^2(\alpha)=e^{-2}\approx 0.135$, which precisely matches the limit
observed in Fig.~\ref{figFmaxN2}.
This behavior points to the asymptotic trend $F_q^{\rm max}(\xi) \sim 0.135 N_{\rm opt}^2$; i.e.\ 
a Fisher information $F_q^{\rm max}(\xi)$ quadratic in the optimal probe size
$N_{\rm opt}$ at small noise $\alpha \rightarrow 1$, reminiscent of the quadratic Fisher 
information $F_q(\xi)=4(1-\kappa)\kappa N^2$ in Eq.~(\ref{Fq2}) at no noise when $\alpha =1$.

\subsection{Optimum at partial entanglement}

It is also relevant to examine the evolution of the quantum Fisher information $F_q(\xi)$ of 
Eq.~(\ref{Fq2g}), at any given size $N$ of the entangled probe, as a function of the Schmidt 
coefficient $\kappa \in [0, 1]$ quantifying the degree of entanglement of the probe from 
Eq.~(\ref{psi0}). It turns out that $F_q(\xi)$ of Eq.~(\ref{Fq2g}) gets maximized at an optimal 
degree of entanglement $\kappa_{\rm opt}$ which is computable analytically as
\begin{equation}
\kappa_{\rm opt} =\dfrac{1}{1+\sqrt{\beta_1 /\beta_0}} \;.
\label{Kopt}
\end{equation}

For the class of unital noises $\mu_0=0$ in Eq.~(\ref{Ns00}), so that $\alpha_0 =\alpha_1$ and 
$\beta_0 =\beta_1$ in Eqs.~(\ref{beta1})--(\ref{beta2}) for any size $N$. For this class of 
noises, it then results from Eq.~(\ref{Kopt}) that $\kappa_{\rm opt}=1/2$, indicating that the 
quantum Fisher information $F_q(\xi)$ of Eq.~(\ref{Fq2g}) is always maximized by a maximally 
entangled probe in Eq.~(\ref{psi0}). By contrast, for nonunital noises with $\mu_0\not =0$ in 
Eq.~(\ref{Ns00}), then in general $\alpha_0 \not =\alpha_1$ and $\beta_0 \not =\beta_1$ in 
Eqs.~(\ref{beta1})--(\ref{beta2}). Depending on the sign of $\mu_0 \in [-1, 1]$, the optimal 
degree of entanglement $\kappa_{\rm opt}$ of Eq.~(\ref{Kopt}) can be found below or above $1/2$. In 
addition, through $\beta_0$ and $\beta_1$ in Eqs.~(\ref{beta1})--(\ref{beta2}), the optimal 
$\kappa_{\rm opt}$ of Eq.~(\ref{Kopt}) is usually dependent on the size $N$ of the probe.
The expression of $\kappa_{\rm opt}(N)$ from Eq.~(\ref{Kopt}) can be placed into Eq.~(\ref{Fq2g}) 
to yield an expression for the Fisher information as $F_q\bigl(\xi; \kappa_{\rm opt}(N) \bigr)$.
This expression can then be maximized according to the size $N$ of the probe; although this can be 
difficult to perform analytically, based on Eqs.~(\ref{Fq2g}) and (\ref{Kopt}) this is quite 
feasible numerically. In this way one obtains the maximizer $N_{\rm opt}$ and therefrom
$(\kappa_{\rm opt}, N_{\rm opt})$ defining the optimal configuration of the probe of 
Eq.~(\ref{psi0}) with size $N_{\rm opt}$ and degree of entanglement $\kappa_{\rm opt}$ maximizing
the quantum Fisher information $F_q(\xi)$ in Eq.~(\ref{Fq2g}). Depending on the values of the
parameters of the nonunital noise, one will usually obtain an optimal probe with finite size 
$N_{\rm opt}$ and partial entanglement $\kappa_{\rm opt}$, representing the most efficient setting 
of the probe of Eq.~(\ref{psi0}) for estimation.

\begin{figure}
{\includegraphics[width=84mm]{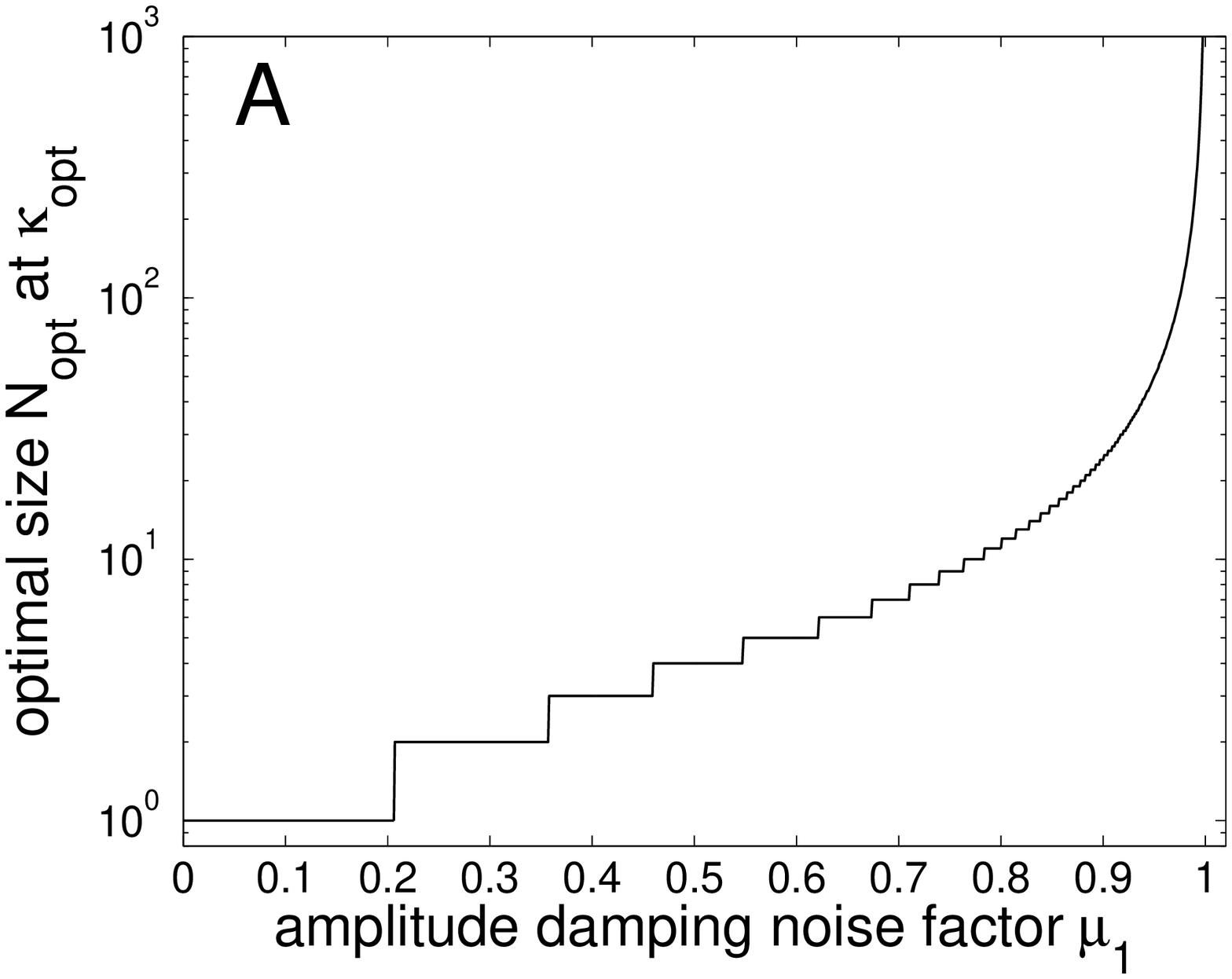}}
\hfill
{\includegraphics[width=84mm]{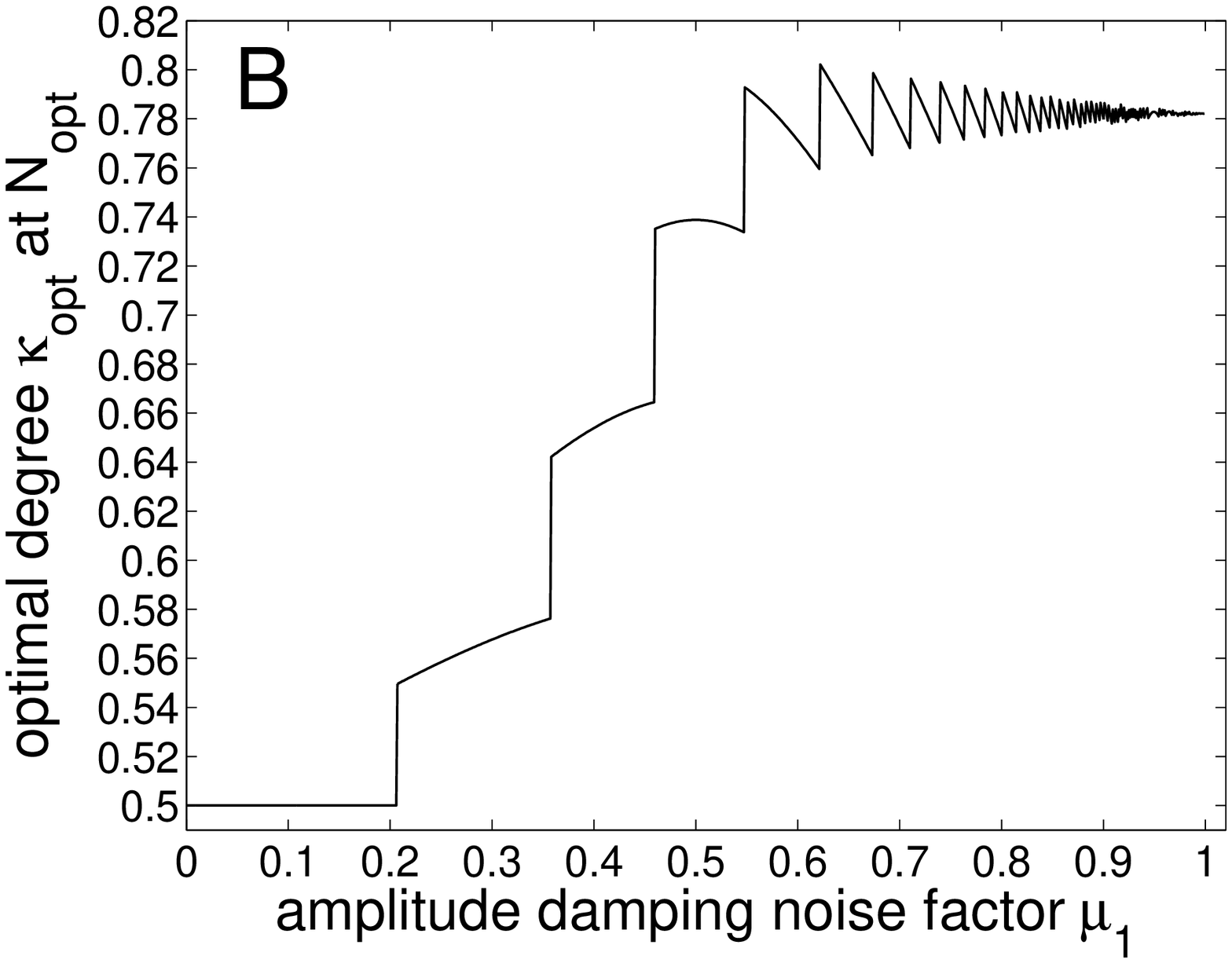}}
\caption[what appears in lof LL p177]
{As a function of the amplitude damping noise factor $\mu_1$, (A) optimal size $N_{\rm opt}$
of the entangled probe, associated with (B) optimal degree of entanglement $\kappa_{\rm opt}$,
together maximizing the quantum Fisher information $F_q(\xi)$ of Eq.~(\ref{Fq2g}).
}
\label{figKopt}
\end{figure}

For illustration we consider the case of the amplitude damping noise \cite{Nielsen00,Wilde13}, 
which describes a qubit asymptotically relaxed to the equilibrium state $\ket{0'}$, and which can 
be obtained in Eqs.~(\ref{Ns00})--(\ref{Ns01}) by $\mu_2=\sqrt{\mu_1}$ and $\mu=1$ so that
$\mu_0=1-\mu_1$. Figure~\ref{figKopt} shows, as a function of the noise factor $\mu_1$, the optimal 
entanglement setting $(\kappa_{\rm opt}, N_{\rm opt})$ maximizing the quantum Fisher information 
$F_q(\xi)$ of Eq.~(\ref{Fq2g}).

\begin{figure}
\centerline{\includegraphics[width=84mm]{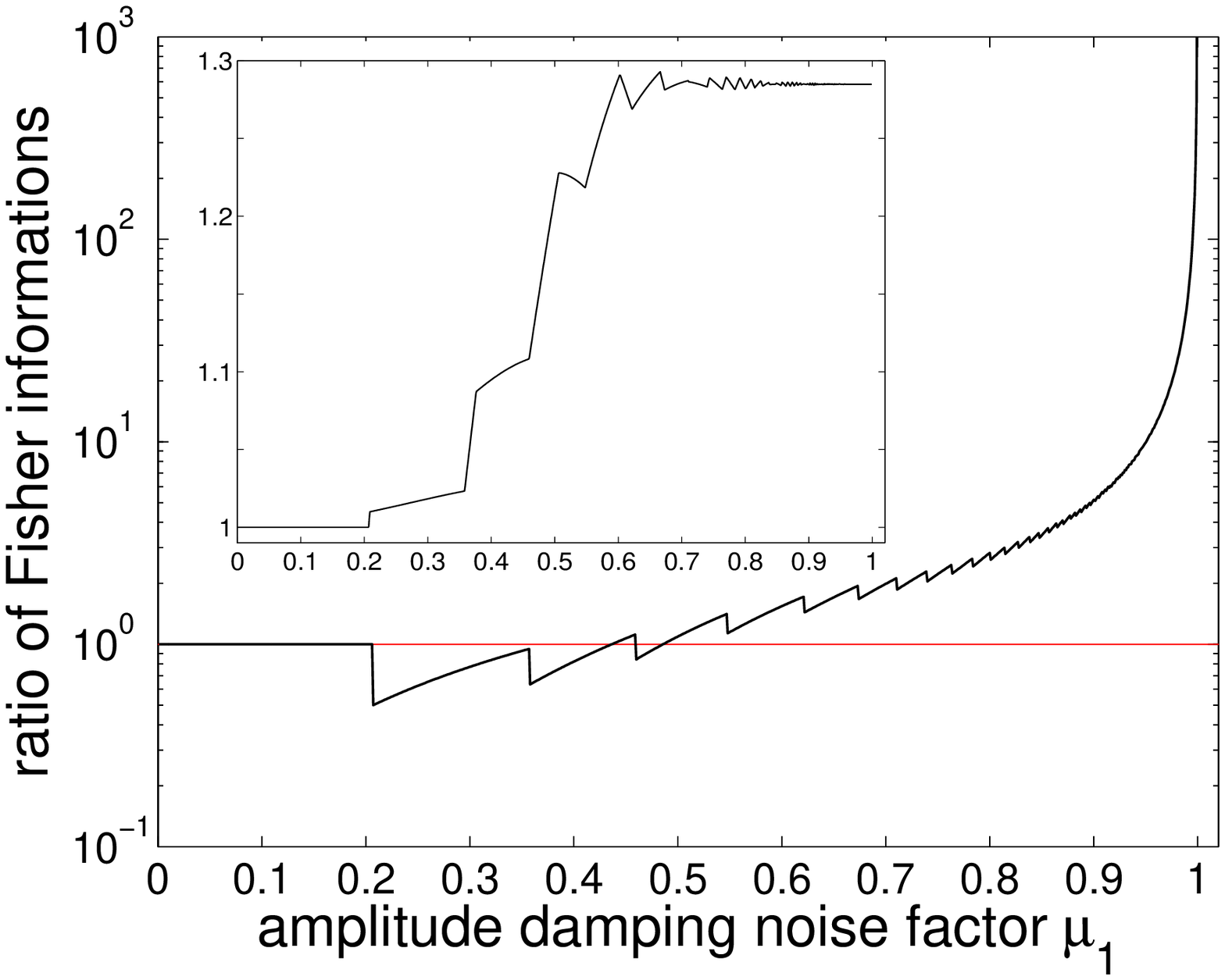}}
\caption[what appears in lof LL p177]
{As a function of the amplitude damping noise factor $\mu_1$, the ratio 
$F_q^{\rm max}(\xi)/(N_{\rm opt} \mu_2^2)$ of the maximum Fisher information $F_q^{\rm max}(\xi)$ 
reached at each $\mu_1$ by $N_{\rm opt}(\mu_1)$ qubits with optimal degree of entanglement 
$\kappa_{\rm opt}(\mu_1)$ as in Fig.~\ref{figKopt}, to the Fisher information $N_{\rm opt} \mu_2^2$ 
reached by the optimal separable probe $\ket{+'}^{\otimes N_{\rm opt}}$ of $N_{\rm opt}$ 
independent qubits. The baseline at ordinate $1$ is shown in red. The inset is the ratio 
$F_q^{\rm max}(\xi ; \kappa=\kappa_{\rm opt})/F_q^{\rm max}(\xi ; \kappa=1/2)$.
}
\label{figFratioAD}
\end{figure}

The optimal entanglement setting $(\kappa_{\rm opt}, N_{\rm opt})$ of Fig.~\ref{figKopt} realizes 
the maximum $F_q^{\rm max}(\xi)$ for the quantum Fisher information $F_q(\xi)$ of Eq.~(\ref{Fq2g}). 
This maximum $F_q^{\rm max}(\xi)$ achieved by $N_{\rm opt}$ optimally entangled qubits, is compared 
in Fig.~\ref{figFratioAD} with the Fisher information $N_{\rm opt} \mu_2^2$ achieved by 
$N_{\rm opt}$ independent qubits prepared in the optimal separable input state 
$\ket{+'}^{\otimes N_{\rm opt}}$.

Figure~\ref{figFratioAD} shows a ratio $F_q^{\rm max}(\xi)/(N_{\rm opt} \mu_2^2)$ which, above 
the noise factor $\mu_1 \approx 0.486$, always exceeds $1$, and which can take very large values 
as the noise vanishes with $\mu_1 \rightarrow 1$ and the size $N_{\rm opt}$ grows. This expresses 
that at intermediate or small noise, the optimally entangled probe is always more efficient than 
the separable probe, with a superiority which can become arbitrary large as $\mu_1 \rightarrow 1$ 
and increasing size $N_{\rm opt}$ of the entangled probe. Moreover, in this range of the noise 
factor $\mu_1$, the optimal 
degree of entanglement $\kappa_{\rm opt}$ is clearly above $1/2$ and lies between $0.66$ and 
$0.81$ in Fig.~\ref{figKopt}(B), indicating that maximum estimation efficiency is achieved at 
partial entanglement. The inset of Fig.~\ref{figFratioAD} depicts the ratio
$F_q^{\rm max}(\xi ; \kappa=\kappa_{\rm opt})/F_q^{\rm max}(\xi ; \kappa=1/2)$ where
$F_q^{\rm max}(\xi ; \kappa=1/2)$ is obtained by maximizing $F_q(\xi)$ at $\kappa=1/2$ instead
of the more efficient maximization at $\kappa=\kappa_{\rm opt}$. Consistently this ratio is
above $1$ and it represents the gain of efficiency by operating at the optimal partial entanglement
$\kappa_{\rm opt}$ instead of operating at maximum entanglement $\kappa=1/2$. The ratio saturates 
around $1.3$ indicating, in these noise conditions, a maximum gain of around $30\%$ by optimizing 
the degree of entanglement. The maximum gain of around $30\%$ is observed in Fig.~\ref{figFratioAD} 
for any noise factor $\mu_1$ above $0.6$. At $\mu_1=0.6$, Fig.~\ref{figKopt}(A) indicates an 
optimal size $N_{\rm opt} =5$ of the entangled probe.

Reference~\cite{Huelga97} tested another family of partially entangled states for frequency 
estimation with phase-flip noise, and it reports improvement of up to $10\%$ compared to the 
maximally entangled probe, in the range of the small probe sizes handled in Ref.~\cite{Huelga97}. 
Here we can handle entangled probes of arbitrary size in Eq.~(\ref{Fq2g}), with maximum improvement 
of around $30\%$ observed in Fig.~\ref{figFratioAD}. Large entangled states with arbitrary degree 
of entanglement are generally difficult to handle analytically (and even numerically), and it is 
useful as here to identify scenarios with entangled states that remain analytically tractable at 
arbitrary sizes, while bringing a meaningful contribution to the picture.
In another context of estimation Refs.~\cite{Ulam01,Giovannetti11} report other situations of 
optimality at partial entanglement of the probe, with a maximum improvement of $\sim 40\%$ above 
the separable probe, while here much larger gain above the separable probe are observed in
Fig.~\ref{figFratioAD}. This indicates the possibility of significant variability of the
quantitative behaviors, depending on the specific conditions, in the presence of noise and 
decoherence which represent sophisticated situations, where intuition can be defeated,
and requiring separate analysis for clarification.

The case of the amplitude damping noise in Figs.~\ref{figKopt}--\ref{figFratioAD} is merely
illustrative, and this important property of optimality at partial entanglement is rather generic 
and occurs in many configurations of nonunital noises inserted in Eq.~(\ref{Fq2g}).
Also we note that when the ratio $F_q^{\rm max}(\xi; N_{\rm opt})/(N_{\rm opt} \mu_2^2)$ of Fisher 
informations with entangled or separable probe, similar to the ratio shown in 
Figs.~\ref{figFratioAD} or \ref{figFratio}, is strictly above $1$, then this is consistently 
associated for the entangled probe with an optimal size $N_{\rm opt} >1$ as in 
Figs.~\ref{figKopt}(A) or \ref{figNopt}(A). In such circumstance, for the asymptotic regime where 
a large number $N$ of qubits are available, an efficient strategy, as also considered for instance 
in Ref.~\cite{Jarzyna13}, would be to group the $N$ qubits into independent blocks of $N_{\rm opt}$ 
optimally entangled qubits. For such strategy with entanglement, the asymptotic regime of large $N$ 
would be characterized, in terms of Fisher information, by an efficiency growing linearly with $N$, 
yet with a level strictly superior to the efficiency of the strategy using the optimal separable 
probe (which also grows linearly with $N$). Therefore, for any conditions with a ratio 
$F_q^{\rm max}(\xi; N_{\rm opt})/(N_{\rm opt} \mu_2^2)$ above $1$, entangled blocks of optimal size 
$N_{\rm opt}$ always improve the asymptotic efficiency above that of optimally prepared independent 
qubits. And the improvement factor can be evaluated, for any noise according to 
Eqs.~(\ref{Ns00})--(\ref{Ns01}), through the analysis of Eq.~(\ref{Fq2g}) as exemplified in this 
Section. This identifies a broad class of conditions with a strategy where entangled probes improve 
over the optimal separable probes for efficient estimation.

\section{With a fraction of inactive qubits}

It is possible to gain additional flexibility in exploiting the $N$-qubit entangled probe of
Eq.~(\ref{psi0}), by letting only a fraction $N_1$ of the $N$ entangled qubits interact with the
unitary $\mathsf{U}_\xi$ plus noise process, while the complementary fraction $N_2=N-N_1$ of
the entangled qubits remain completely untouched in the operation. In practice this could 
represent a situation where, after preparing the $N$-qubit entangled probe, $N_2$ qubits are kept 
sheltered in the laboratory while $N_1$ qubits are sent outside to interact with the noise and 
unitary process to be estimated. Comparable schemes with inactive entangled quantum systems or 
ancillas have also been considered for various scenarios of quantum metrology, with no noise 
\cite{Martini03,DAriano03,Giovannetti04,Giovannetti06}, or with noise 
\cite{Giovannetti11,Demkowicz14,Smirne15}, often with two subsystems -- active and inactive -- 
of equal size. A prior motivation here is that the active qubits interact both with the process to 
be estimated (which is useful) and with the noise (which is detrimental), so a mixed strategy 
involving some entangled inactive qubits may globally lead to a more favorable trade-off (as we 
are going to verify in some configurations).

From the initial $N$-qubit probe $\ket{\psi_0}$ of Eq.~(\ref{psi0}), with $N_1$ active qubits
we obtain the transformed state
$\ket{\psi_1}=\mathsf{U}_\xi^{\otimes N_1} \otimes \mathrm{I}_2^{\otimes N_2} \ket{\psi_0} = 
\sqrt{1-\kappa}\, e^{-iN_1\xi/2} \ket{0'_N} + \sqrt{\kappa}\, e^{iN_1\xi/2}\ket{1'_N}$, with 
density operator
$\rho_1=\ket{\psi_1}\bra{\psi_1}$ as
\begin{equation}
\rho_1 (\xi)=(1-\kappa) \ket{0'_N}\bra{0'_N} + \kappa\ket{1'_N}\bra{1'_N} + 
\sqrt{(1-\kappa)\kappa}\, 
\Bigl( e^{iN_1\xi} \ket{1'_N}\bra{0'_N} + e^{-iN_1\xi} \ket{0'_N}\bra{1'_N} \Bigr)
\label{ro1red}
\end{equation}
instead of Eq.~(\ref{ro1}).
The noisy state is $\rho_\xi=\mathcal{N}^{\otimes N_1}\otimes \mathcal{I}^{\otimes N_2}(\rho_1)$,
with $\mathcal{I}(\cdot)$ the identity superoperator for the qubit, yielding
\begin{eqnarray}
\nonumber
\rho_\xi &=&(1-\kappa) \mathcal{N}^{\otimes N_1}\otimes \mathcal{I}^{\otimes N_2}
\bigl( \ket{0'_N}\bra{0'_N} \bigr) + \kappa
\mathcal{N}^{\otimes N_1}\otimes \mathcal{I}^{\otimes N_2} \bigl( \ket{1'_N}\bra{1'_N} \bigr) \\
&& \mbox{} + \sqrt{(1-\kappa)\kappa}\; \Bigl( 
\alpha_2^{*N_1} e^{iN_1\xi} \ket{1'_N}\bra{0'_N} + 
\alpha_2^{N_1} e^{-iN_1\xi} \ket{0'_N}\bra{1'_N} \Bigr) 
\label{ro_xired}
\end{eqnarray}
replacing Eq.~(\ref{ro_xi}). Next, we have the noisy basis state 
\begin{eqnarray}
\nonumber
&& \mathcal{N}^{\otimes N_1}\otimes \mathcal{I}^{\otimes N_2} \bigl( \ket{0'_N}\bra{0'_N} \bigr) =
\biggl[ \dfrac{1}{2}(1+\alpha_0)\ket{0'}\bra{0'}+\dfrac{1}{2}(1-\alpha_0)\ket{1'}\bra{1'} \biggr]
^{\otimes N_1} \otimes \bigl(\ket{0'}\bra{0'} \bigr)^{\otimes N_2} \\
&& \qquad \mbox{} = \sum_{\vec{u} \in \{0', 1' \}^{N_1}} 
\biggl[ \dfrac{1}{2}(1+\alpha_0)\biggr]^{N_1-w(\vec{u})} 
\biggl[ \dfrac{1}{2}(1-\alpha_0)\biggr]^{w(\vec{u})}
\ket{\vec{u}}\bra{\vec{u}} \otimes \bigl(\ket{0'}\bra{0'} \bigr)^{\otimes N_2}
\label{N00_red} 
\end{eqnarray}
in place of Eq.~(\ref{N00}), and a comparable expression for 
$\mathcal{N}^{\otimes N_1}\otimes \mathcal{I}^{\otimes N_2} \bigl( \ket{1'_N}\bra{1'_N} \bigr)$
replacing Eq.~(\ref{N11}). By the same approach, we obtain as before an explicit
characterization of both $\rho_\xi$ and $\partial_\xi \rho_\xi$. For the eigendecomposition
of $\rho_\xi$ the structure of the eigenstates is conserved, with only two eigenstates
belonging to the plane $\bigl(\ket{0'_N}, \ket{1'_N} \bigr)$ to impact the Fisher information
$F_q(\xi)$, and determined via the matrix elements
\begin{eqnarray}
\label{ro_00r}
\rho_{00} &=& (1-\kappa) \biggl[ \dfrac{1}{2}(1+\alpha_0)\biggr]^{N_1} +
                 \kappa  \biggl[ \dfrac{1}{2}(1-\alpha_1)\biggr]^{N_1} \delta_{NN_1} \;, \\ 
\label{ro_11r}
\rho_{11} &=&    \kappa  \biggl[ \dfrac{1}{2}(1+\alpha_1)\biggr]^{N_1} +
              (1-\kappa) \biggl[ \dfrac{1}{2}(1-\alpha_0)\biggr]^{N_1} \delta_{NN_1} \;, \\ 
\label{ro_01r}
\rho_{01} &=& \sqrt{(1-\kappa)\kappa}\, \alpha_2^{N_1}  e^{-iN_1\xi} \;,
\end{eqnarray}
incorporating the Kronecker delta $\delta_{NN_1}$ as an essential evolution relative to
Eqs.~(\ref{ro_00})--(\ref{ro_01}). The two associated eigenvalues sum according to
$\lambda_+ + \lambda_-=\rho_{00}+\rho_{11}$ as in Eq.~(\ref{lam1_lam2}),
with now the two parameters
\begin{eqnarray}
\label{beta1b}
\beta_0 &=& \biggl[ \dfrac{1}{2}(1+\alpha_0)\biggr]^{N_1} +
            \biggl[ \dfrac{1}{2}(1-\alpha_0)\biggr]^{N_1} \delta_{NN_1} \;, \\
\label{beta2b}
\beta_1 &=& \biggl[ \dfrac{1}{2}(1+\alpha_1)\biggr]^{N_1} +
            \biggl[ \dfrac{1}{2}(1-\alpha_1)\biggr]^{N_1} \delta_{NN_1} \;.
\end{eqnarray}
Through the same path, the
quantum Fisher information $F_q(\xi)$ of Eq.~(\ref{Fq1}) finally evaluates to
\begin{equation}
F_q(\xi) = 
\dfrac{4(1-\kappa)\kappa N_1^2 \mu_2^{2N_1}}{(1-\kappa)\beta_0 + \kappa\beta_1 } \;.
\label{Fq2r}
\end{equation}
which coincides with Eq.~(\ref{Fq2g}) only when $N_1=N$.

The Fisher information of Eq.~(\ref{Fq2r}) indicates that the efficiency obtained with a given 
fraction $N_1$ of active qubits, is independent of the total size $N$ and is a function of $N_1$ 
only, provided $N \ge N_1 +1$. In this respect, as soon as the size $N=N_1+1$ can the efficiency 
of Eq.~(\ref{Fq2r}) be obtained, i.e.\ with only $N_2=1$ inactive qubits. For $N_1$ given, there 
is no need to entangle more than $N=N_1 +1$ qubits to obtain the efficiency of Eq.~(\ref{Fq2r}). 
The presence of $N_2=1$ inactive qubit does affect the efficiency, but $N_2=2$ or more inactive 
qubits do not add anything except the burden of having to process a larger size $N$ overloaded 
with superfluous inactive qubits.

In given noise configurations, due to the similarity of form with Eq.~(\ref{Fq2g}), the quantum 
Fisher information $F_q(\xi)$ in Eq.~(\ref{Fq2r}) also returns exponentially to zero at large 
$N_1 \rightarrow \infty$, and is maximized at an optimum of the size $N_1$ and at an optimal 
degree of entanglement $\kappa_{\rm opt}$. Especially, this $\kappa_{\rm opt}$ is still given by 
Eq.~(\ref{Kopt}) and is generally different from $1/2$ marking maximal entanglement.

\begin{figure}[htb]
\centerline{\includegraphics[width=84mm]{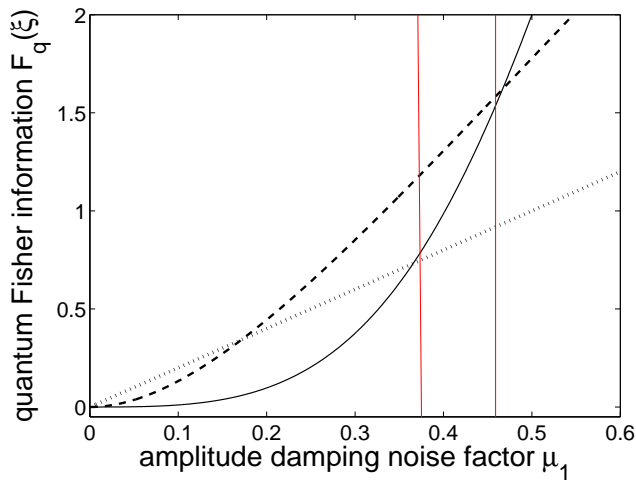}}
\caption[what appears in lof LL p177]
{As a function of the amplitude damping noise factor $\mu_1$, for a total probe size $N=3$, the 
quantum Fisher information $F_q(\xi)$ of Eq.~(\ref{Fq2g}) with all $N$ entangled qubits active 
(solid line), $F_q(\xi)$ of Eq.~(\ref{Fq2r}) with $N_1=N-1$ active qubits among the $N$ entangled 
qubits (dashed line), and $F_q(\xi)=N_1 \mu_2^2$ for $N_1$ independent qubits optimally prepared 
in the separable state $\ket{+'}^{\otimes N_1}$ (dotted line).
Between the two vertical red lines at $\mu_1=0.357$ and $\mu_1=0.459$ is the region of $\mu_1$ 
where the size $N=3=N_{\rm opt}$ is the optimum for the probe with all $N$ active qubits as
indicated in Fig.~\ref{figKopt}(A).
}
\label{figN1}
\end{figure}

To illustrate the benefit of handling a reduced fraction $N_1 <N$ of active qubits, instead of 
the totality of $N$ active qubits, we again turn to the case of the amplitude damping noise as in 
Figs.~\ref{figKopt}--\ref{figFratioAD}. Figure~\ref{figN1} compares the quantum Fisher 
information $F_q(\xi)$ achieved by an $N$-qubit entangled probe with the totality of $N$ active 
qubits, or with a reduced fraction $N_1=N-1$ of active qubits. In each scenario, the $N$-qubit
entangled probe is prepared at the optimal degree of entanglement $\kappa_{\rm opt}$ of 
Eq.~(\ref{Kopt}) as it follows either from Eqs.~(\ref{beta1})--(\ref{beta2}) or from
Eqs.~(\ref{beta1b})--(\ref{beta2b}).

Figure~\ref{figN1} shows a situation where a probe with a reduced fraction $N_1=N-1$ of active 
qubits exhibits a Fisher information $F_q(\xi)$ surpassing that of the probe with the totality 
$N$ of active qubits, this occurring for sufficiently large noise levels when $\mu_1 < 0.467$.
A reduced fraction $N_1=N-1$ of active qubits can also surpass the Fisher information of $N_1$
independent active qubits optimally prepared in the separable state $\ket{+'}^{\otimes N_1}$. 
Figure~\ref{figN1} shows that, for a noise factor $\mu_1 \in [0.172, 0.467]$, the 
setting with a reduced fraction $N_1=N-1$ of active qubits surpasses the two other settings of the
probe. In addition, the size $N=3$ addressed in Fig.~\ref{figN1} is the optimal size $N_{\rm opt}$ 
for the probe with all $N$ active qubits when the noise factor $\mu_1 \in [0.357, 0.459]$, as 
indicated in Fig.~\ref{figKopt}(A). Over this range of $\mu_1$, where the probe with the reduced 
fraction $N_1$ is the best probe, it surpasses the probe with all $N$ active qubits when it is at 
its best. This represents a proof of principle that, in definite noise configurations, 
accommodating an inactive qubit in the $N$-qubit entangled probe, can stand as the most efficient 
setting for estimation. 
We emphasize that this inactive qubit must be coherently prepared as part of the entangled probe
in Eq.~(\ref{psi0}), and coherently measured with it.
Instead of being measured, if this inactive qubit were discarded, the effect would be
described by tracing out this qubit in the state $\rho_\xi$ of Eq.~(\ref{ro_xired}). The 
consequence would be to cancel in $\rho_\xi$ the term carrying the dependence with $\xi$, and to 
produce in this way a reduced state independent of $\xi$, as a result of the randomization
provoked by discarding the inactive qubit. And such a reduced state independent of $\xi$ would of
course be of no use for estimating $\xi$ and would entail a zero Fisher information $F_q(\xi)$.
To obtain the performance of Fig.~\ref{figN1}, it is therefore essential to measure the inactive 
qubit coherently with the entangled probe, and as we have seen, although this qubit never interacts 
with the process $\mathsf{U}_\xi$ being estimated, its presence can improve the estimation 
efficiency. This is a manifestation of counterintuitive properties that can result from quantum 
entanglement, and here bearing useful significance for estimation.

Here also the case of the amplitude damping noise presented in 
Fig.~\ref{figN1} is merely illustrative, and analysis of Eq.~(\ref{Fq2r}) shows that this 
property of maximum efficiency with an inactive qubit in the probe is generic, and exists in other 
noise configurations. For instance it exists with the depolarizing noise of 
Figs.~\ref{figFq1}--\ref{figFmaxN2}, over some range of the noise factor $\alpha$.

In general, the most efficient strategy for estimating the qubit phase $\xi$, will depend on 
the specific conditions. In particular, it will depend on whether one has access to a fixed budget 
$N$ of entangled qubits, corresponding to the maximum size one is able to manage for the probe, or 
if on the contrary one is able to adjust the size $N$ at the optimal level $N_{\rm opt}$ as in 
Fig.~\ref{figNopt}(A) or \ref{figKopt}(A) irrespective of its (high) value. It will also depend on 
the type and level of the noise, knowing that, qualitatively, as also reflected in other reports 
\cite{Huelga97,Giovannetti11}, separable states are usually preferable at large noise while 
entangled states become superior at intermediate or at small noise levels. In any case, the 
optimal strategy, both in terms of probe size and degree of entanglement, can 
be worked out in given conditions by maximizing the quantum Fisher information $F_q(\xi)$ which 
is explicitly known via Eqs.~(\ref{Fq2g}) or (\ref{Fq2r}), and confronting to $\mu_2^2$ per 
independent qubit each prepared in state $\ket{+'}$; and such a strategy is applicable to any 
axis $\vec{n}$ associated with the qubit rotation by the unknown phase angle $\xi$.

\section{Discussion}

The present study considered a generic estimation task consisting in estimating the phase $\xi$ 
of a unitary transformation $\mathsf{U}_\xi$ on a qubit, equivalent in Bloch representation to a
rotation by $\xi$ around an arbitrary axis $\vec{n}$. The effect of quantum noise or decoherence 
was included under the form of the general noise model of Eqs.~(\ref{Ns00})--(\ref{Ns01}), which 
by commuting with $\mathsf{U}_\xi$ has the ability to 
represent as a single lumped operation decohering actions which could take place anywhere over the 
whole process starting with preparation of the probe up to its measurement. The noise model of 
Eqs.~(\ref{Ns00})--(\ref{Ns01}) includes as special cases important decohering processes such as 
depolarizing noise, phase-flip noise, amplitude and generalized amplitude damping noise describing 
the interaction of the qubit with an environment represented by a thermal bath at arbitrary 
temperature \cite{Nielsen00,Gillard18}. For estimation of the phase $\xi$ in the presence of noise, 
the quantum Fisher information $F_q(\xi)$ was used to assess the overall best achievable 
efficiency especially governing the mean-squared estimation error.

An optimal separable input probe was characterized as the pure qubit state $\ket{+'}$ matched to 
the arbitrary rotation axis $\vec{n}$, along with the maximum Fisher information $\mu_2^2$ per 
qubit it achieves for the general noise model of Eqs.~(\ref{Ns00})--(\ref{Ns01}).
The possibility of enhanced efficiency was then investigated by employing an $N$-qubit
entangled probe prepared in the family of states $\ket{\psi_0}$ of Eq.~(\ref{psi0}) accommodating 
an arbitrary degree of entanglement $\kappa$.
In such general conditions, an analytical expression was derived with Eq.~(\ref{Fq2g}) 
for the quantum Fisher information $F_q(\xi)$.

This expression of Eq.~(\ref{Fq2g}) shows that in the absence of noise, the $N$-qubit entangled 
probe of Eq.~(\ref{psi0}) always displays the superefficient scaling as $N^2$ of the Fisher 
information $F_q(\xi)$, for any non-degenerate degree of entanglement $\kappa$, and is in this 
respect superior to the Fisher information $N$ achieved by the $N$-quibt optimal separable probe.
This is the superefficiency property afforded by entanglement in noise-free conditions, also 
previously reported in other situations, and found here to hold across a whole family of 
partially entangled probes. Moreover, in the presence of noise or decoherence according to the 
model of Eqs.~(\ref{Ns00})--(\ref{Ns01}), the expression of Eq.~(\ref{Fq2g}) shows that,
for all (non-vanishing) noise configurations, the Fisher information $F_q(\xi)$ always returns to 
zero asymptotically at large size $N$ of the entangled probe. In addition, there always exists a 
finite size $N_{\rm opt}$ of the entangled probe, along with an optimal degree of entanglement 
$\kappa_{\rm opt}$, together maximizing the Fisher information $F_q(\xi)$ at a level 
$F_q^{\rm max}(\xi)$ which is completely characterized, depending 
on the specific noise configuration. We established that for all unital noises (with $\mu_0 =0$) in 
Eqs.~(\ref{Ns00})--(\ref{Ns01}), the optimal degree of entanglement is $\kappa_{\rm opt} =1/2$, 
i.e.\ at maximum entanglement of the probe in Eq.~(\ref{psi0}). By contrast, for nonunital noises 
in Eqs.~(\ref{Ns00})--(\ref{Ns01}), optimality is often obtained at partial entanglement, for 
$\kappa_{\rm opt} \not \in \{0, 1/2, 1\}$. The maximum $F_q^{\rm max}(\xi)$ achieved by 
$(N_{\rm opt}, \kappa_{\rm opt})$ in Eq.~(\ref{Fq2g}), can be confronted with the maximum Fisher 
information $N_{\rm opt}\mu_2^2$ achieved by the $N_{\rm opt}$-quibt optimal separable probe, and 
is usually found superior across a significant range of noise conditions, essentially at 
intermediate or small noise levels. In such conditions, grouping the $N$ qubits into independent 
blocks formed of $N_{\rm opt}$ entangled qubits, restores in the asymptotic regime of large $N$, 
a nonvanishing efficiency improving over that of $N$ independent qubits optimally prepared.

The effect of letting only a reduced fraction $N_1$ among $N$ entangled qubits in the probe 
interact with the process $\mathsf{U}_\xi$ plus noise was also investigated, by means of 
Eq.~(\ref{Fq2r}) which provides an explicit evaluation of the quantum Fisher information 
$F_q(\xi)$ in these conditions. It revealed that arranging for one inactive qubit (but more 
than one is unnecessary) in the entangled probe, sometimes stands as the most efficient setting 
for estimation. Such inactive qubit, although it never interacts with the process 
$\mathsf{U}_\xi$ being estimated, can nevertheless improve the estimation efficiency.

Owing to the importance of entanglement and decoherence for quantum estimation, several other 
important studies such as Refs.~\cite{Huelga97,Shaji07,Ji08,Ma11,Tesio11,Smirne15} have dealt with 
the subject. However, our derivations of Eqs.~(\ref{Fq2g}) and (\ref{Fq2r}) with their general 
conditions of applicability provide new elements here, especially with their analysis pointing to 
optimal probe size and degree of entanglement, according to the configurations over a broad class of 
quantum noises. Other studies often focused on frequency rather than phase estimation, especially 
to optimize the interaction time of the probe with the process being estimated, and with no optimal 
probe size $N_{\rm opt}$ and degree of entanglement $\kappa_{\rm opt}$ reported. 
Here instead, we consider 
phase estimation with a fixed interaction time of the probe with the process being estimated, such 
as for instance the fixed traveling time of photons across an interferometer, so as to produce the 
phase shift $\xi$. And we exhibit optimal conditions for the probe size $N_{\rm opt}$ and 
degree of entanglement $\kappa_{\rm opt}$. 
As long as we can relate the phase $\xi$ as $\xi=\nu T$ to 
an angular frequency $\nu$ acting over a fixed interaction time $T$, the associated Fisher 
informations can be related as $F_q(\nu)=T^2 F_q(\xi)$. Therefore, with a fixed time $T$, the 
evolutions and optimizations according to $N$ and $\kappa$ concerning $F_q(\xi)$ of 
Eq.~(\ref{Fq2g}) or Eq.~(\ref{Fq2r}) as reported here, apply equally to $F_q(\nu)=T^2 F_q(\xi)$, 
and consequently the present characterizations hold for phase as well as for frequency estimation.
In these respects our results stand as useful complements, for a broader appreciation 
of the role of entanglement for quantum parameter estimation and quantum metrology, and for 
fuller understanding and mastering of entanglement for the benefit of information processing.


\end{document}